\documentclass[10pt,journal]{IEEEtran}

\ifCLASSINFOpdf
\else
\fi
\usepackage{array,booktabs}
\usepackage{algpseudocode}
\usepackage{algorithm}
\usepackage{amsfonts}
\usepackage{amsmath}
\usepackage{amssymb}
\usepackage{amsthm}
\usepackage{graphicx}
\usepackage[noadjust]{cite}
\usepackage{mathrsfs}
\usepackage{stmaryrd}
\usepackage{siunitx}
\usepackage{multicol}
\usepackage{tikz}
\usepackage{setspace}
\usepackage{pstricks, pst-node, pst-plot, pst-circ}
\usepackage{moredefs}
\usetikzlibrary{shapes}

\makeatletter
\newtheoremstyle{newcorollary}
  {3pt}
  {3pt}
  {}
  {1em}
  {\itshape}
  {:}
  {.5em}
  {\thmname{#1}\thmnumber{\@ifnotempty{#1}{ }#2}%
   \thmnote{ {\the\thm@notefont(\itshape#3)}}}
\makeatother
\theoremstyle{newcorollary}

\makeatletter
\newtheoremstyle{newdefinition}
  {3pt}
  {3pt}
  {}
  {1em}
  {\itshape}
  {:}
  {.5em}
  {\thmname{#1}\thmnumber{\@ifnotempty{#1}{ }#2}%
   \thmnote{ {\the\thm@notefont(\itshape#3)}}}
\makeatother
\theoremstyle{newdefinition}
\newtheorem*{definition}{Definition}

\makeatletter
\newtheoremstyle{newlemma}
  {3pt}
  {3pt}
  {}
  {1em}
  {\itshape}
  {:}
  {.5em}
  {\thmname{#1}\thmnumber{\@ifnotempty{#1}{ }#2}%
   \thmnote{ {\the\thm@notefont(\itshape#3)}}}
\makeatother
\theoremstyle{newlemma}
\newtheorem{lemma}{Lemma}

\makeatletter
\newtheoremstyle{newtheorem}
  {3pt}
  {3pt}
  {}
  {1em}
  {\itshape}
  {:}
  {.5em}
  {\thmname{#1}\thmnumber{\@ifnotempty{#1}{ }#2}%
   \thmnote{ {\the\thm@notefont(\itshape#3)}}}
\makeatother
\theoremstyle{newtheorem}
\newtheorem{theorem}{Theorem}

\makeatletter
\newtheoremstyle{newproposition}
  {3pt}
  {3pt}
  {}
  {1em}
  {\itshape}
  {:}
  {.5em}
  {\thmname{#1}\thmnumber{\@ifnotempty{#1}{ }#2}%
   \thmnote{ {\the\thm@notefont(\itshape#3)}}}
\makeatother
\theoremstyle{newproposition}

\makeatletter
\newtheoremstyle{newproof}
  {3pt}
  {3pt}
  {}
  {2em}
  {\itshape}
  {:}
  {.5em}
  {\thmname{#1}\thmnumber{\@ifnotempty{#1}{ }#2}%
   \thmnote{ {\the\thm@notefont(\itshape#3)}}}
\makeatother
\theoremstyle{newproof}
\newtheorem*{proofs}{Proof}

\makeatletter
\newtheoremstyle{newremark}
  {3pt}
  {3pt}
  {}
  {2em}
  {\itshape}
  {:}
  {.5em}
  {\thmname{#1}\thmnumber{\@ifnotempty{#1}{ }#2}%
   \thmnote{ {\the\thm@notefont(\itshape#3)}}}
\makeatother
\theoremstyle{newproof}
\newtheorem*{remark}{Remark}

\providelength{\AxesLineWidth}       
\providelength{\plotwidth}           
\providelength{\LineWidth}           
\providelength{\MarkerSize}          
\newrgbcolor{GridColor}{0.8 0.8 0.8}%

\begin{document}
\title{Optimality of Rate Balancing in Wireless\\Sensor Networks}
\author{Alla~Tarighati,~\IEEEmembership{Student Member,~IEEE,}
	 and~Joakim~Jald{\'e}n,~\IEEEmembership{Senior Member,~IEEE,}

\thanks{The authors are with the ACCESS Linnaeus Centre, department of signal processing,
KTH Royal Institute of Technology, Stockholm 100 44, Sweden (e-mail: allat@kth.se; jalden@kth.se).
%
}
\thanks{The material in this paper was presented in part at the 16th IEEE Int. workshop signal
  process. advances in wireless commun. (SPAWC), Stockholm, Sweden 2015.
}}
\maketitle


\begin{abstract}

We consider the problem of distributed binary hypothesis testing in a parallel network topology where sensors independently observe some phenomenon and send a finite rate summary of their observations to a fusion center for the final decision. We explicitly consider a scenario under which (integer) rate messages are sent over an error free multiple access channel, modeled by a sum rate constraint at the fusion center. This problem was previously studied by Chamberland and Veeravalli, who provided sufficient conditions for the optimality of one bit sensor messages. Their result is however crucially dependent on the feasibility of having as many one bit sensors as the (integer) sum rate constraint of the multiple access channel, an assumption that can often not be satisfied in practice. This prompts us to consider the case of an a-priori limited number of sensors and we provide sufficient condition under which having no two sensors with rate difference more than one bit, so called \emph{rate balancing}, is an optimal strategy with respect to the Bhattacharyya distance between the hypotheses at the input to the fusion center. We further discuss explicit observation models under which these sufficient conditions are satisfied.
\end{abstract}

\IEEEpeerreviewmaketitle

\section{Introduction}\label{sec:intro}
A central problem in wireless sensor networks is that of decentralized detection, where spatially separated sensors receive information about the state of some phenomenon and send summaries of their observations to a fusion center (FC) over rate constrained channels. The fusion center then makes a final decision about the state of the phenomenon based on the aggregate information received from the remote sensors \cite{Veer12,Cham03,Cham07,li07}. A large body of research in decentralized detection has been devoted to the case where the sensors transmit their information to the FC through parallel access channels, commonly known as the parallel topology \cite{Varsh96}. However, in wireless sensor networks the wireless medium is typically shared among the sensors, and the sensor to FC channels are arguably more reasonably modeled as a common multiple access channel (MAC). 
This setting was previously studied by Chamberland and Veeravalli \cite{Cham03} who judiciously argued that when an unbounded number of senors with independent observations compete for rate under a sum rate constraint at the input of the FC, it is often optimal to use as many sensors as possible and let each sensor communicate with the FC over a one bit link. The purpose of our work is to in the same vein study when equal rate allocation, or rate balancing, is an optimal solution for a fixed number of sensors operating under a common sum rate constraint. It is worth stressing here that the MAC channel model used herein, as in \cite{Cham03}, is a set of error free channels that are subject to a sum rate constraint. A number of recent works \cite{Berg09,Li11,Ciu13} consider distributed detection over, arguably more realistic, wireless MAC channels where the main focus is on fading, interference, and channel state information. These works are also different from our work in that the main focus is on fusion rules, typically assuming binary signaling, while our focus is mainly on the design and rates of the multi-rate sensor rules.

To be specific,
we consider a binary hypothesis testing problem where the phenomenon or hypothesis $H$ is from the set $\mathcal{H}\triangleq\{0,1\}$ and where the observations at the $N$ remote sensors are independent and identically distributed conditioned on the true hypothesis $H$. Sensor $n$, where $n=1,\ldots,N$, is required to quantize its own observation into an $r_n$ bit message where\footnote{Herein, we let $\mathbb{N} \triangleq \{0, 1, 2, \ldots \}$ denote the set of natural numbers including $0$.} $r_n \in \mathbb{N}$ before transmission to the FC, and we assume that $\sum_{n=1}^N r_n \leq R$ for some $R \in \mathbb{N}$. Chamberland and Veeravalli \cite{Cham03} studied the structure of an optimal sensor configuration in this scenario in terms of the optimal number of sensors $N$ and the optimal set of rates $r_n$ for $n=1,\ldots,N$. They proposed the Chernoff information at the input of the FC as a measure of optimality given the intractability of the Bayesian probability of error as a design criteria \cite{Poor77}, and proved that having $N=R$ one bit (binary) sensors is optimal if there exists a one bit sensor rule with a Chernoff information of the sensor output that is at least half the Chernoff information of the original observation. Moreover, they proved the existence of such a sensor rule when the observations are drawn from particular Gaussian and exponential observation models.

Although the condition of \cite{Cham03} leads to a very simple network design, involving $R$ identical one bit sensors implemented as simple likelihood ratio tests, it is in many cases simply not practically feasible to have an arbitrary number of sensors. Hardware or spatial constraints will often limit the maximum number of sensors deployed in practice. We therefore wish to extend \cite{Cham03} and consider the case where $N$ is fixed or limited a-priori, and consider the problem of optimally selecting the set of rates $\{r_n \}_{n=1}^N$. When $N \geq R$ the problem of selecting the rates and designing the sensors rules again reverts to the problem studied in \cite{Cham03}, but it remains open for the case where $N < R$. However, unlike \cite{Cham03} we will consider the Bhattacharyya distance between the conditional distributions at the FC input, referred to as the joint sensor index space \cite{Lon90}. The main reason for using the Bhattacharyya distance in place of the Chernoff information is that it will increase the tractability of the problem, and allow us to handle the added difficulties of considering higher rate sensors. This said, it should however be noted that the Bhattacharyya distance: 1) has been frequently used in the past as a performance measure in the design of distributed detection systems \cite{Poor77,Lon90}; 2) provides a lower bound on the Chernoff information; and 3) provides an upper bound on the Bayesian probability of error at the FC. The main contribution of our work is to provide sufficient conditions under which a balanced rate allocation is optimal in the sense that it provides maximal Bhattacharyya distance over all rate allocations and over all possible sensor decision rules. We then prove that the sufficient conditions are satisfied in the case of known signals in additive Laplacian noise.  We further conjecture that rate balancing is optimal also for the case of known signals in additive Gaussian noise, and provide compelling supporting evidence for this conjecture, although we have so far been unable to prove this stringently. Both the Laplacian and Gaussian noise models are common to the distributed detection literature, see for example \cite{Swa93, Sun01, Ald04}. The key implication of our work is that it can for many commonly considered observation models be judiciously argued that wireless sensor networks should ideally be symmetrically designed in terms of the sensors and their communications rates.

The outline of the paper is as follows. We formalize the problem in Section \ref{sec:prob}, and provide some prerequisite definitions and results, restate the result of 
Chamberland and Veeravalli \cite{Cham03} in order to introduce notation and make the comparison between the two results self contained. We then provide our central result in Section \ref{sec:res} by providing sufficient conditions under which rate balancing is an optimal strategy. We follow this by studying the obtained conditions in the explicit cases of signal in additive Laplacian and Gaussian noise in Section \ref{sec:applic}, and illustrate the applicability of the results in a simple sensor design problem in Section \ref{sec:disc}. Finally, we conclude the work in Section \ref{sec:conc}.

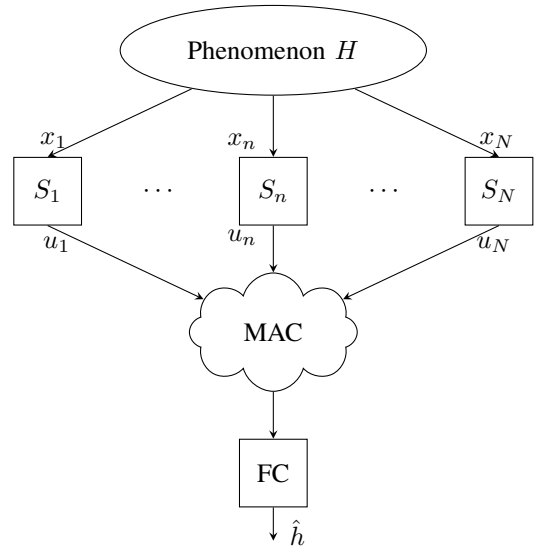
\begin{figure}
\centering
\begin{tikzpicture}[align=center,scale=0.75,>=stealth] 
\node (S1) at (-4,3) [rectangle,minimum size=0.9cm,draw] {$S_1$};
\node (dots1) at (-2,3) {$\cdots$};
\node (Sn) at (0,3) [rectangle,minimum size=0.9cm,draw] {$S_n$};
\node (dots2) at (2,3) {$\cdots$};
\node (SN) at (4,3) [rectangle,minimum size=0.9cm,draw] {$S_N$};
\node (PH) at (0,5.5) [ellipse,inner sep=3mm,draw] {Phenomenon $H$};
\draw [->] (PH) -- (S1.north) node  [near end,left,inner sep=6pt] {$x_{1}$};
\draw [->] (PH) -- (Sn.north) node  [near end,left,inner sep=6pt] {$x_{n}$};
\draw [->] (PH) -- (SN.north) node  [near end,right,inner sep=6pt] {$x_{N}$};
\node (MAC) at (0,0.5) [cloud, draw,cloud puffs=8,cloud puff arc=150, aspect=2, inner ysep=.7em] {MAC};
\draw [->] (S1.south) -- (MAC) node  [near start,left,inner sep=6pt] {$u_{1}$};
\draw [->] (Sn.south) -- (MAC) node  [near start,left,inner sep=6pt] {$u_{n}$};
\draw [->] (SN.south) -- (MAC) node  [near start,right,inner sep=6pt] {$u_{N}$};
\node (FC) at (0,-2) [rectangle,minimum size=0.9cm,draw] {FC};
\draw [->] (MAC) -- (FC) node  [near start,right,inner sep=6pt] {};
\draw [->] (FC) -- (0,-3.2) node  [near end,right,inner sep=6pt] {{$\hat{h}$}};
\end{tikzpicture} 
\caption{Setting of the sensors in a network, where the sensors send their data through a MAC channel to the FC.}
\label{fig:topology}
\end{figure}
\section{Problem Statement}\label{sec:prob}
As noted in the introduction, we consider a binary hypothesis testing problem with $N$ sensors $S_1,\ldots,S_N$ arranged as shown in Fig.~\ref{fig:topology}. 
Sensor $S_n$, $n=1,\ldots,N$, makes an observation $x_{n}$, computes a message $u_{n}$, and sends this message towards the FC. We assume that the observations are conditionally independent given the true hypothesis, and let $X_n \sim X$ and $U_n \sim U$ be random variables corresponding to observation and message of sensor $S_n$, respectively. The phenomenon $H$ is modeled as a random variable drawn from $\mathcal{H}\triangleq\{0,1\}$ with a prior probabilities $\pi_0$ and $\pi_1$, respectively, and gives rise to conditionally independent and identically distributed observations $X_{n} \in \mathcal{X}$ for $n=1,\ldots,N$ with conditional distribution $f_{X|H}(x|h)$ for $h \in \{ 0,1 \}$ at the sensors. 

Given a realization $x_{n}$ of $X_{n}$, sensor $S_n$ computes its $r_n$ bit message $u_{n} \in \mathcal{U}_{r_n} \triangleq \{ 1, \ldots, 2^{r_n}  \}$, where $r_n \in \mathbb{N}$, using a (measurable) decision function $\gamma_n:\mathcal{X} \to \mathcal{U}_{r_n}$, i.e.,
\begin{equation}
\gamma_n(x_{n})=u_{n} \, .
\label{eq:gammadef}
\end{equation} 
The FC makes the final decision $\hat{h}\in\mathcal{H}$ based on the aggregate sensor messages using a (measurable) decision function $\gamma_0: \underline{\mathcal{U}}\to \mathcal{H}$, i.e., \[\gamma_0(u_{1},\ldots,u_{N})=\hat{h}\,,\]where $\underline{\mathcal{U}}\triangleq \mathcal{U}_{r_1}\times\ldots\times\mathcal{U}_{r_N}$. As in \cite{Cham03} we model the MAC channel as a joint constraint on the information rate from the sensors to the FC, i.e.,
\begin{equation}
\sum_{n=1}^N r_n\leq R \, ,
\label{eq:Rconstraint}
\end{equation}
and assume that communication errors (from sensors to the FC) are negligible. In other words, each sensor $S_n$ is capable of sending its message with a maximum rate of $r_n$  integer bits reliably to the FC. The overall objective is to design the decision functions $\gamma_n$, for $n=0,\ldots,N$, of the FC and the sensors, and to allocate rate to the sensors in such a way that some measure of performance is optimized. As noted in the introduction, this MAC channel is very different from the MAC channels considered in, e.g., \cite{Berg09,Li11,Ciu13}. The channel in those works corresponds to simultaneous transmission over a common shared resource, while the channel considered herein and in \cite{Cham03} is more akin to communication over a shared and limited number of orthogonal resource blocks.

Given a sensor rule $\gamma_n$ for sensor $S_n$ and the true hypothesis $h \in \mathcal{H}$, the probability mass function (pmf) associated with the message $U_{n} = \gamma_n(X_{n}) \in \mathcal{U}_{r_n}$ can be obtained as
\begin{equation}\begin{split}
P_{U_{n}\vert H}(u\vert h)&=\Pr\left\{\gamma_n(X_{n})=u \vert h\right\}\\
&=\int_{x \in \gamma_n^{-1}(u)}\!f_{X \vert H}(x \vert h)\,dx \,,
\label{eq:singlePMF}\end{split}\end{equation}
where $\gamma_n^{-1}(u)$ is the set of observations $x \in \mathcal{X}$ that satisfy $\gamma_n(x)=u$. The conditional pmf of the message vector $\underline{U} \triangleq (U_{1},\ldots,U_{N})$ can be obtained using the observation vector $\underline{X} \triangleq (X_{1}\ldots,X_{N})$ according to
\begin{equation}\begin{split}
P_{\underline{U} \vert H}\left(\underline{u}\vert h\right)&=\Pr\left\{ \left(\gamma_1(X_{1}),\ldots,\gamma_N(X_{N})\,\right)=\underline{u}\vert h\right\}\\
&=P_{U_{1}\vert H}(u_1\vert h)\ldots P_{U_{N}\vert H}(u_N\vert h)\\
&=\prod_{n=1}^{N}\int_{x \in \gamma_n^{-1}(u_n)}\!f_{X \vert H}(x \vert h)\,dx\,,
\label{eq:multiplePMF}\end{split}\end{equation}
where $\underline{u}\triangleq (u_1,\ldots,u_N)$, and where the second line follows by the independence of the observations $X_{n}$ and thus the different sensor messages $U_{n}$. 

Given a rate allocation $\underline{r} \triangleq (r_1,\ldots,r_N)$ and a set of sensor rules $\underline{\gamma} \triangleq (\gamma_1,\ldots,\gamma_N)$, it is well known that the Bayesian error probability at the FC, $P_\mathrm{E} \triangleq \Pr(\hat{H}\neq H)$, is minimized when the FC applies the maximum a-posteriori (MAP) rule, and that the MAP rule can be implemented as a likelihood ratio test applied to $P_{\underline{U} \vert H}\left(\underline{u}\vert h\right)$. It has however also been widely acknowledged that the probability of error criteria does not in general lead to tractable design procedures for $\underline{\gamma}$, which has led authors to consider other measures of performance such as Ali-Silvey distance measures \cite{Poor77} applied to $\underline{U}$, or measures such as the Chernoff information \cite{Tsi88}.

The \emph{Chernoff information} at the input of the FC can, for a given rate allocation $\underline{r}$ and set of sensor rules $\underline{\gamma}$ be defined\footnote{The Chernoff information (or Chernoff exponent) \cite{Cher52} can also be defined as the negative rate of decay of the MAP error probability over repeated  observations \cite{Cham03, Tsi88}, which is in fact how it was defined in \cite{Cham03}. The two definitions lead to the same results, and we choose to use \eqref{eq:chernoff} as the definition only because it simplifies exposition.} using $P_{\underline{U}\vert H}\left(\underline{u}\vert h\right)$ according to
\begin{equation}\begin{split}
&\mathcal{C}_{\underline{r}}\left(\,\underline{\gamma}\,\right)\triangleq \\&-\min_{0\leq \alpha \leq 1}\log \left[ \sum_{\underline{u}\in \underline{\mathcal{U}}} \left[P_{\underline{U}\vert H}\left(\underline{u}\vert 0\right)\right]^\alpha\left[P_{\underline{U}\vert H}\left(\underline{u}\vert 1\right)\right]^{1-\alpha}\right ]\,.
\label{eq:chernoff}\end{split}\end{equation}
One can similarly define the Chernoff information delivered by a single sensor according to
\begin{equation}\begin{split}
&\mathcal{C}_{r_n}\!(\gamma_n) \triangleq \\&-\!\min_{0\leq \alpha \leq 1}\!\log\! \Bigg[ \sum_{u \in \mathcal{U}_{r_n}}\! \left[P_{{U_n}\vert H}\!\left({u}\vert 0\right)\right]^\alpha
\left[P_{{U_n}\vert H}\!\left({u}\vert 1\right)\right]^{1-\alpha}\!\Bigg]\,,
\label{eq:Cb}\end{split}\end{equation}
and the Chernoff information of a single observation as
\begin{equation}\begin{split}
\mathcal{C}&_\infty=\\&-\!\min_{0\leq \alpha \leq 1}\!\log \int_{\mathcal{X}}\!\! \left[f_{X\vert H}\!\left(x\vert 0\right)\right]^\alpha\left[f_{X\vert H}\!\left(x\vert 1\right)\right]^{1-\alpha}dx \, .
\label{eq:Cinf}\end{split}\end{equation}
For any $r_n \in \mathbb{N}$ and $\gamma_n$ it can be shown that $\mathcal{C}_{r_n}(\gamma_n) \leq \mathcal{C}_\infty$ \cite{Cham03}, and $$\lim_{r_n \rightarrow \infty}\mathcal{C}_{r_n}\!\left(\gamma_n\right) = \mathcal{C}_\infty$$ under some mild technical conditions for most reasonable sensor designs. Based on the independence of the observations it can also be shown that
\begin{equation}
\mathcal{C}_{\underline{r}}\left(\,\underline{\gamma}\,\right) \leq \sum_{n=1}^N \mathcal{C}_{r_n}\!\left(\gamma_n\right)\,.
\label{eq:Csum}
\end{equation}
The inequality is however not in general tight given that the optimal $\alpha$ in \eqref{eq:Cb} may depend on $n$ and not coincide with the optimal $\alpha$ in \eqref{eq:chernoff}. Thus, it is not in general possible to express $\mathcal{C}_{\underline{r}}\left(\,\underline{\gamma}\,\right)$ as the sum of the Chernoff information $\mathcal{C}_{r_n}\!\left(\gamma_n\right) $ delivered by each sensor.

Chamberland and Veeravalli considered the problem of maximizing the Chernoff information $\mathcal{C}_{\underline{r}}(\,\underline{\gamma}\,)$, as defined in \eqref{eq:chernoff}, over $N$, $\underline{r}$, and $\underline{\gamma} \in \underline{\Gamma}_{\underline{r}}$ where $\underline{\Gamma}_{\underline{r}} = \Gamma_{r_1} \times \cdots \times \Gamma_{r_N}$ and $\Gamma_r$ is the set of all possible rate $r$ sensor decision functions $\gamma:\mathcal{X} \to \mathcal{U}_{r}$ \cite{Cham03}. Albeit simpler than minimizing the Bayesian probability of error $P_\mathrm{E}$, this problem is still very hard \cite{Tsi93}, especially as the optimization of the joint set of sensor rules $\underline{\gamma}$ over $\underline{\Gamma}_{\underline{r}}$ does not in general for any given $\underline{r}$ decouple into separate optimization problems over $\Gamma_{r_n}$ for each $\gamma_n$ for $n=1,\ldots,N$. However, they were able to formulate a sufficient condition for when it is optimal to choose $N=R$ and $r_1 = \cdots = r_N = 1$, i.e., to only use one bit (binary) sensor decisions, and established that this condition was satisfied for some relevant distributions. For completeness, we paraphrase their main result below.

\begin{theorem}[Chamberland and Veeravalli] \label{prop:CH}
Suppose that there exists a binary (one bit) function $\gamma : \mathcal{X} \mapsto \{1,2\}$ for which\footnote{It should be mentioned that in the original paper \cite{Cham03} the message set $\{0,1\}$ was used, while here without loss of generality and because of consistency with the definition of $\mathcal{U}_{r_n}$, we use the message set $\{1,2\}$.}
\begin{equation}
2\,\mathcal{C}_{1}\!\left(\gamma\right)\geq\mathcal{C}_\infty\,,
\label{eq:CHresult}\end{equation}
then having $N=R$ identical sensors, each sending a single bit of information, is optimal.
\end{theorem}

Given the problem formulation, the proof is in retrospect straightforward, building on that $\mathcal{C}_{r_n}\!\left(\gamma_n\right) \leq \mathcal{C}_\infty$ together with \eqref{eq:CHresult} implies that the sum of the Chernoff information of two optimal single bit sensors is larger than (or as large as) the Chernoff information of a single sensor of any rate, along with the observation that \eqref{eq:Csum} holds with equality if all (optimized) $\gamma_n$ are identical which can be assumed if $r_1 = \cdots = r_N = 1$. In short, one can improve upon any given high rate design by replacing any sensor of rate $r_n$ with $r_n$ optimum rate-one sensors without decreasing the Chernoff information $\mathcal{C}_{\underline{r}}\left(\,\underline{\gamma}\,\right)$ at the input of the FC or violating the sum rate constraint \eqref{eq:Rconstraint}.

Although the restriction to identical one bit sensors greatly simplifies the network design, it is not always a feasible strategy when there is a limit to the maximum number of active sensors. Having $N$ rate-one sensors cannot in general be optimal when, for instance, the maximum number of active sensors $N$ is less than the MAC channel rate $R$. To see this, one can consider a network of $N-1$ optimal rate-one sensors and include an optimal sensor of rate $R-N+1$ bits. This latter strategy will by construction satisfy the rate constraint \eqref{eq:Rconstraint}, and will outperforms a network using only single bit sensors as long as $\mathcal{C}_{r_n}\!\left(\gamma_n\right)$ is strictly increasing in $r_n$ which it usually is.

The main contribution of our work is to derive a sufficient condition for when rate balancing is an optimal strategy in the sense that one can without loss of generality assume that the rate of any two sensors differs by at most one bit, i.e., when
\begin{equation}
r_{\mathrm{max}} - r_{\mathrm{min}} \leq 1 \quad \text{and} \quad
\sum_{n=1}^{N}r_n=R \,,
\label{eq:ratebalance}
\end{equation}
where $r_{\mathrm{max}}$ and $r_{\mathrm{min}}$ are the highest and the lowest allocated rates, respectively.
However, a problem with extending the argument of \cite{Cham03} to general rate allocations is that we cannot in general assume that \eqref{eq:Csum} provides a tight bound. In particular, the optimizing $\alpha$ may differ between \eqref{eq:chernoff}, \eqref{eq:Cb}, and \eqref{eq:Cinf}, and may in \eqref{eq:Cb} depend on the rate $r_n$.\footnote{There are also additional technical difficulties in the proof of Theorem~\ref{th:bstar} that makes continuation with the Chernoff information difficult.}
In order to circumvent this difficulty we will replace the optimization over $\alpha$ in both \eqref{eq:chernoff} and \eqref{eq:Cb}, regardless of $r_n$, by setting $\alpha = 0.5$. This reduces the Chernoff information to the Bhattacharyya distance \cite{Kai67}. Although this may seem completely ad-hoc, it is worth noting that the Bhattacharyya distance -- a member of the class of Ali-Silvey distances \cite{Poor77} -- has previously been used in its own right as a design criteria for quantizer design in decentralized hypothesis testing \cite{Poor77,Lon90}.

The Bhattacharyya distance associated with a rate allocation ${\underline{r}}$ and a set of sensor decision functions $\underline{\gamma}$ is thus given by
\begin{equation}
\mathcal{B}_{\underline{r}}\left(\,\underline{\gamma}\,\right)\triangleq -\log\left[ \sum_{\underline{u}\in \underline{\mathcal{U}}} \sqrt{P_{\underline{U}\vert H}\left(\underline{u}\vert 0\right)\,P_{\underline{U}\vert H}\left(\underline{u}\vert 1\right)}\,\right ]\,.
\label{eq:bhatta}\end{equation}
In the same manner as for the Chernoff information, we define the Bhattacharyya distance of a single sensor $S_n$ with rate $r_n$ and decision function $\gamma_n$ as
\begin{equation}
\mathcal{B}_{r_n}\!\left(\gamma_n\right)\triangleq -\log\left[ \sum_{u\in \mathcal{U}_{r_n}} \sqrt{P_{{U_n}\vert H}\left({u}\vert 0\right)\,P_{{U_n}\vert H}\left({u}\vert 1\right)}\,\right ]\, , 
\label{eq:bhattsingle}\end{equation}
and the Bhattacharyya distance of a single observation as
\begin{equation}
\mathcal{B}_\infty \triangleq -\log\left[ \int_{\mathcal{X}} \sqrt{f_{X\vert H}\left({x}\vert 0\right)\,f_{X\vert H}\left({x}\vert 1\right)} \, dx \,\right ]
\, .
\label{eq:bhattinf}\end{equation}
It immediately follows that $\mathcal{B}_{\underline{r}}\left(\,\underline{\gamma}\,\right) \leq \mathcal{C}_{\underline{r}}\left(\,\underline{\gamma}\,\right)$, $\mathcal{B}_{r_n}\!\left(\gamma_n\right) \leq \mathcal{C}_{r_n}\!\left(\gamma_n\right)$, and $\mathcal{B}_\infty \leq \mathcal{C}_\infty$ as the Bhattacharyya distances can be obtained from \eqref{eq:chernoff}, \eqref{eq:Cb} and \eqref{eq:Cinf} with $\alpha = 0.5$ in place of the optimization over $\alpha$ as noted above. The Bhattacharyya distance can also be shown to provide an upper bound on the Bayesian probability of error of the MAP FC detector according to $P_\mathrm{E} \leq \sqrt{\pi_0 \pi_1} e^{-\mathcal{B}_{\underline{r}}(\underline{\gamma})}$ \cite{Boekee79}. The benefit (in terms of mathematical tractability) of considering the Bhattacharyya distance instead of the Chernoff information is captured by the following, easily proven, lemma.

\begin{lemma}The Bhattacharyya distance of a network of sensors, arranged as in Fig.~\ref{fig:topology} and with independent observations, is equal to sum of the Bhattacharyya distances of individual sensors, i.e.,
\begin{equation*}\begin{split}
\mathcal{B}_{\underline{r}}\left(\,\underline{\gamma}\,\right)
&=\sum_{n=1}^N\mathcal{B}_{r_n}\!\left(\gamma_n\right)\,.
\end{split}\end{equation*}
\label{lem:bhatt}\end{lemma}
\begin{proofs}
See Appendix \ref{app:lemm1}.
\end{proofs}
What Lemma \ref{lem:bhatt} implies is that the Bhattacharyya distances $\mathcal{B}_{r_n}\!\left(\gamma_n\right)$ of the single sensors for $n=1,\ldots,N,$ completely describe the Bhattacharyya distance $\mathcal{B}_{\underline{r}}\left(\,\underline{\gamma}\,\right)$ of the network. Thus, a network which maximizes the Bhattacharyya distance at the FC is a network with individually optimized sensors for any fixed rate allocation $\underline{r}$, since
\[\max_{\underline{\gamma} \in \underline{\Gamma}_{\underline{r}}}\mathcal{B}_{\underline{r}}\left(\,\underline{\gamma}\,\right)=\sum_{n=1}^{N}\max_{\gamma_n \in \Gamma_{r_n}}\mathcal{B}_{r_n}\!\left(\gamma_n\right)\,.\]
This significantly simplifies the overall problem that we consider. To simplify the problem of designing (or optimizing) each individual sensor, let
\begin{equation}
l(x) \triangleq \ln \frac{f_{X\vert H}(x\vert 1)}{f_{X\vert H}(x\vert 0)}
\label{eq:llr}
\end{equation}
be the log-likelihood ratio for a given observation $x \in \mathcal{X}$, let $l_n = l(x_n)$, let $L_n = l(X_n)$, and let $f_{L|H}(l \vert h)$ be the conditional distribution of $L_n$ induced by $f_{X|H}(x|h)$ and \eqref{eq:llr}. 
Let $t_0\triangleq -\infty$, $t_K\triangleq +\infty$, and $t_1,\ldots,t_{K-1} \in \mathbb{R}$ be a set of thresholds that satisfy $t_0\leq t_1 \leq \ldots \leq t_{K-1}\leq t_K$ for some $K \in \mathbb{N}$ where $K \geq 1$, and let $\mathcal{I}_1\triangleq[t_0,t_1], \, \mathcal{I}_2\triangleq[t_1,t_2],\ldots,\,\mathcal{I}_i\triangleq[t_{i-1},t_{i}],\ldots,\,\mathcal{I}_{K}\triangleq[t_{K-1},t_K]$ be a set of $K$ intervals defined by the thresholds, so that $\mathbb{R} = \cup_{i=1}^{K} \mathcal{I}_i$ with overlap only at the interval boundaries. 
We say that a sensor decision function $\gamma_n$ is a \emph{monotone likelihood quantizer} if $u = \gamma_n(x)$ implies that $l(x) \in \mathcal{I}_u$. The central result of \cite{Tsi93Ext} is that it can without loss of generality be assumed that the decision functions that maximize  $\mathcal{B}_{r_n}\!\left(\gamma_n\right)$ over $\gamma_n \in \Gamma_{r_n}$ are monotone likelihood quantizers, i.e., there is a monotone likelihood quantizer $\gamma_n^\star \in \Gamma_{r_n}$ for which $\mathcal{B}_{r_n}\!\left(\gamma_n\right) \leq \mathcal{B}_{r_n}\!\left(\gamma_n^\star\right)$ for all $\gamma_n \in \Gamma_{r_n}$. Thus, the problem of designing sensor $S_n$ effectively reduces to selecting $K-1$ log-likelihood thresholds where $K=2^{r_n}$ and where $r_n$ is the rate allocated to sensor $S_n$. The same claim of optimality of monotone likelihood quantizers can also be made regarding the optimization of $\mathcal{C}_{r_n}\!\left(\gamma_n\right)$ over $\gamma_n \in \Gamma_{r_n}$ \cite{Tsi93Ext}, but as noted this does in and of itself immediately imply that $\underline{\gamma}$ is optimized over $\underline{\Gamma}_{\underline{\gamma}}$ by a set of monotone likelihood quantizers.\footnote{Note that the assumption that the communications channels are error free plays a crucial role in this argument. For error prone channels the performance would in general also depend on the interval to message mapping used, while for error free channels one can assume an arbitrary interval to message mapping.}

Furthermore, when the log-likelihood ratio $l(x)$ of the observations at the sensors is monotone in $x \in \mathcal{X} \subseteq \mathbb{R}$, one can without loss of generality also assume that a monotone quantizer is applied directly to the observations $x_n$ rather than to the log-likelihood values.
One example of this, that we will also consider later, is when the observation model is a (conditionally) known signal in additive Laplacian noise with scale parameter $s$, i.e., where the observations at the sensors are distributed according to
\begin{equation}
f_{X \vert H}(x \vert h) = \frac{1}{2s}\,e^{-\frac{\vert {x-m_h}\vert}{s}}
\label{eq:Laplacase}\end{equation}
where $m_h$ is a hypothesis dependent mean. The Laplacian noise distribution is often used in practice as a generic model of heavy-tailed noise. Another example, also considered later and undoubtedly the most commonly considered noise model, is when the noise is zero mean Gaussian with variance $\sigma^2$, and where the observations at the sensors are distributed according to
\begin{equation}\begin{split}
f_{X \vert H}(x\vert h) = \frac{1}{\sqrt{2\pi \sigma^2}}\,e^{-\frac{({x-m_h})^2}{2\sigma^2}} \, .
\label{eq:Gausscase}\end{split}\end{equation}
As the set of achievable $\mathcal{C}_{r_n}\!\left(\gamma_n\right)$ and $\mathcal{B}_{r_n}\!\left(\gamma_n\right)$ are not affected by invertible transformations of the observations we may without loss of generality assume that $m_0 = -m$ and $m_1=m$, and that  $s=1$ and $\sigma^2=1$ in the Laplacian case and the Gaussian case, respectively, and we will do so in what follows. The problem of designing an optimal sensor decision rule $\gamma_n$ with observation distribution according to \eqref{eq:Laplacase} or \eqref{eq:Gausscase} and with rate $r_n$ can thus without loss of optimality be reduced to the problem of selecting $2^{r_n}-1$ thresholds, or $2^{r_n}$ intervals $\mathcal{I}_1,\ldots,\mathcal{I}_{2^{r_n}}$, for the observation $x_n$. For clarity of exposition, we will for this reason from now on restrict attention to the case where $x \in \mathbb{R}$ and only explicitly consider monotone quantizers for which $u = \gamma_n(x)$ implies $x \in \mathcal{I}_u$. Although this does not uniquely identify $u = \gamma_n(x)$ when $x = t_k$ for some $k \in \{ 1, \ldots, K \}$, i.e., when the observation falls on the border of an interval, this will not affect our results as we only explicitly consider observation models without point-masses in $f_{X \vert H}(x\vert h)$ which implies that $X \neq t_k$ almost surely. The extension to the case of observation models with point masses can be straightforwardly handled by the selection of (deterministic) rules for breaking ambiguities \cite{Tsi93Ext}, but we will omit the explicit treatment of this in order not to obscure our main points. Our results will hold also for observation models with non-monotone log-likelihoods and more complex observation spaces $\mathcal{X}$ after replacing $x$ by $l(x) \in \mathbb{R}$.

While the remaining sensor design problem is still non-trivial, these simplifications do introduce enough structure to formulate verifiable sufficient conditions under which rate balancing is a provably optimal strategy. Section \ref{sec:res} builds up to the main result of this paper, namely a sufficient condition for the optimality of rate balancing in the same flavor as the one of Theorem \ref{prop:CH}.

\section{Main Results}\label{sec:res}
We begin by introducing the notion of concavity for discrete functions. Theorem \ref{th:main1} then establishes the partial result that concavity of $\mathcal{B}_{r_n}\!\left(\gamma_n\right)$ in $r_n$ for the optimally designed $\gamma_n$ is sufficient for optimality of rate balancing in the sense of \eqref{eq:ratebalance}.

\begin{definition}
A function $g:\mathbb{N}\to \mathbb{R}$ is a discrete concave function over $\mathbb{N}$ if \cite{Mur03}
\begin{equation}g(r-1)+g(r+1)\leq 2g(r)\,, \quad \forall r \in \mathbb{N} \, , \, r > 0 \,.\label{eq:concdef}\end{equation}
\end{definition}
The following lemma follows straightforwardly for any discrete concave function $g(r)$ by iteratively using the definition above, and is given without proof. 
\begin{lemma}
For any discrete concave function $g(r)$ it holds that
\begin{equation}\begin{split}
g(r+k)+g(r-k)&\leq 2g(r)\,,\\
g(r+k+1)+g(r-k)&\leq g(r+1)+g(r)\,,
\end{split}\end{equation}
for all $r,k \in \mathbb{N}$, where $r\geq k$.
\label{lem:concavity}\end{lemma}

Now, let $\Gamma_r$ be the set of all possible decision rules for a sensor $S_n$ at rate $r_n=r$. Let $\gamma_n^\star$ be a decision rule which maximizes the Bhattacharyya distance, and let $\mathcal{B}_r^\star$ be this maximum Bhattacharyya distance, i.e.,
\[\gamma_n^\star=\arg\max_{\gamma\in\Gamma_r}\,\{\mathcal{B}_r\!\left(\gamma\right)\}\,,\]
and
\[\mathcal{B}_r^\star\triangleq\mathcal{B}_r\!\left(\gamma_n^\star\right)\,.\]
If, for a given observation distribution at the sensors, $\mathcal{B}_r^\star$ is a discrete concave function in the rate $r$, then it follows from Lemma \ref{lem:concavity} that two rate balanced (and individually optimized) sensors dominates any other pair of two sensors. More precisely, one can replace two sensors of rates $r+k$ and $r-k$ by two sensors that each has rate $r$ without reducing $\mathcal{B}_{\underline{r}}\left(\,\underline{\gamma}\,\right)$, and one can replace two sensors of rates $r+k+1$ and $r-k$ by two minimum difference sensors of rates $r$ and $r+1$ without reducing $\mathcal{B}_{\underline{r}}\left(\,\underline{\gamma}\,\right)$. This is formalized by the following theorem. 

\begin{theorem}
If, for a given observation distribution at the sensors, $\mathcal{B}_r^\star$ is a discrete concave function in the rate $r$, then rate balancing in the sense of \eqref{eq:ratebalance} is an optimal rate allocation.
\label{th:main1}\end{theorem}
\begin{proofs}
Consider a network of $N$ sensors, $S_1,\ldots,S_N$, with rate allocation $\underline{r} \triangleq (r_1,\ldots,r_N)$. We can without loss of generality assume\footnote{This assumption can always be achieved by simply relabeling the sensors if necessary.} that $r_1\leq r_2\leq\ldots\leq r_N$. Assume that the sensors use a set of optimal sensor decision functions $\underline{\gamma}^\star = (\gamma^\star_1,\ldots,\gamma^\star_N)$ for the rate allocation $\underline{r}$, in the sense that $\mathcal{B}_{\underline{r}}\left(\,\underline{\gamma}\,\right) \leq \mathcal{B}_{\underline{r}}\left(\,\underline{\gamma}^\star\,\right)$ for all $\underline{\gamma} \in \Gamma_{r_1} \times \cdots \times \Gamma_{r_N}$. If $\mathcal{B}_r^\star$ is concave, it follows by Lemma \ref{lem:concavity}, we can replace sensors $S_1$ and $S_N$ with two rate balanced sensors, say $S'_1$ and $S'_N$, with rates $r'_1$ and $r'_N$ that satisfy
\begin{equation} (r'_1,r'_N) = \left\{ \begin{array}{ll}
          (r,r) & \mbox{if $r_1+r_N=2r$}\vspace{2mm}\\
(r,r+1)  & \mbox{if $r_1+r_N=2r+1$}\,, \end{array}\right.
\label{eq:ratepair} \end{equation}
which implies that $r_1\leq r'_1\leq r'_N\leq r_N$ and $r_1'+r_N' = r_1+r_N$, and decision functions ${\gamma_1^\star}'$ and ${\gamma_N^\star}'$ that satisfy $\mathcal{B}_{r_1}^\star + \mathcal{B}_{r_N}^\star \leq\mathcal{B}_{r'_1}^\star + \mathcal{B}_{r'_N}^\star$. 
By additionally letting $S_n' = S_n$, ${\gamma_n^\star}' = \gamma_n^\star$ and $r_n' = r_n$ for $n=2,\ldots,N-1$ we obtain a new rate allocation $\underline{r}' \triangleq (r_1',\ldots,r_N')$ and set of decision functions ${\underline{\gamma}^\star}' = ({\gamma_1^\star}',\ldots,{\gamma_N^\star}')$ for which $\mathcal{B}_{\underline{r}}\left(\,\underline{\gamma}^\star\,\right) \leq \mathcal{B}_{\underline{r}'}\left(\,{\underline{\gamma}^\star}'\,\right)$ and $\sum_{n=1}^N r_n' = \sum_{n=1}^N r_n$. We can in the same way iteratively replace the lowest-rate sensor and the highest-rate sensor in the new network with two rate balanced sensors without decreasing the Bhattacharyya distance, until we have a rate balanced sensor network, i.e., until there are no two sensors in the network with a rate difference more than one. This establishes that rate balancing is an optimal rate allocation strategy. \qed
\end{proofs}

\begin{remark}
The result of Theorem \ref{th:main1} can also be obtained by appealing to the Schur-concavity \cite{Mar10} of $\mathcal{B}^\star_{\underline{r}} = \sum_{n=1}^N \mathcal{B}^\star_{r_n}$ in $\underline{r}$. This connection is further explored in Section \ref{sec:disc} in relation to a comparison of different rate allocations.
\end{remark}

Theorem \ref{th:main1} establishes that concavity of $\mathcal{B}_r^\star$ is sufficient for the optimality of rate balancing. This said, obtaining the optimal $\gamma_n^\star$ with respect to either $\mathcal{B}_{r_n}\!\left(\gamma_n\right)$ or $\mathcal{C}_{r_n}\!\left(\gamma_n\right)$ is still hard, 
although there do exist numerical optimization procedures at least capable of achieving locally optimal designs \cite{Ald04}, and it will be difficult to analytically characterize $\mathcal{B}_r^\star$ in general. However, some intuitive support for the concavity of $\mathcal{B}_r^\star$ can be obtained from prior work on high rate quantization \cite{Poor88,Ben89}. In particular, Benitz and Bucklew proposed asymptotically optimal quantization rules for which the Bhattacharyya distance and Chernoff information can be (asymptotically) obtained in closed form \cite{Ben89}. The idea behind this design method is to uniformly quantize the interval $[0,1]$ and let a \emph{companding function} $q:\mathcal{X}\to [0,1]$ define the quantization of $\mathcal{X}$ by mapping the uniform thresholds over $[0,1]$ to thresholds in $\mathcal{X}$ through $q^{-1}$. The (asymptotically) optimal companding function $q$ depends on the conditional distributions at the sensors $f_{X\vert H}(x\vert h)$, and the key result in \cite{Ben89} is a set of conditions that define the asymptotically optimal $q$ in terms of the Chernoff information in the high rate regime, i.e., when $r\to\infty$. However, using the asymptotically optimal companding function $q$ together with a finite rate quantization of $[0,1]$ has been empirically observed to work well also for finite rates $r$, see \cite{Alla15}. Further, the quantizers constructed using this methodology for the observation models in \eqref{eq:Laplacase} and \eqref{eq:Gausscase} are symmetric in a way that implies that the optimizing $\alpha$ in the definition of the Chernoff information in \eqref{eq:chernoff} and \eqref{eq:Cb} is given by $\alpha = 0.5$. Thus, the Bhattacharyya distance and Chernoff information coincide for these designs.

Following the procedure outlined in \cite{Ben89} for the Laplacian observation model in \eqref{eq:Laplacase}, the asymptotically optimal companding function $q$ can be shown to equal
\[q(x)=\left\{ \begin{array}{ll}
        0 &\quad \quad\quad x< -m\,,\\
        \frac{x}{2m}+\frac{1}{2} & -m \leq x\leq m\,,\\
                  1 &\quad\quad\quad x> m\, .\end{array} \right.\]
For the Gaussian observation model in \eqref{eq:Gausscase} the asymptotically optimal companding function $q$ is given by
\[q(x)=1-\mathcal{Q}\left(\frac{x}{\sqrt{3}} \right)\,,\]
where $\mathcal{Q}(y)$ is the tail probability of unit-variance Gaussian density defined as
\begin{equation}
\mathcal{Q}(y) \triangleq\int_y^\infty\!\frac{1}{\sqrt{2\pi}}e^{-\frac{\,\,t^2}{2\,}}\,dt\,.
\label{eq:Qfun}
\end{equation}
The resulting Bhattacharyya distances of a rate-$r$ sensor become $\mathcal{B}_r=\beta_r+ o\!\left(2^{-2r}\right)$, where
\begin{equation}
\beta_r=m-\log\left[1+m+\frac{m^3}{6}\,2^{-2r} \right]
\label{eq:BenBhattLapla}\end{equation}
for the Laplacian case, and where
\begin{equation}
\beta_r=\frac{m^2}{2}-\log\left[1+\frac{\pi\sqrt{3}\,m^2}{4}\,2^{-2r} \right]
\label{eq:BenBhattGauss}\end{equation}
for the Gaussian case.
At high rates, the term $o\!\left(2^{-2r}\right)$ vanishes, and it can be shown from first principles that for both the aforementioned cases
\[\beta_{r-1}+\beta_{r+1}\leq 2\beta_{r}\,,\]
i.e., the asymptotic Bhattacharyya distances of sensors designed using this method are \emph{asymptotically} discrete concave functions of the rate $r$. Combined with Theorem \ref{th:main1} this plausibly suggests that for a large $R$, it is optimal to have a rate balanced network of high rate sensors which are designed according to Benitz and Bucklew's methodology. However, due to the uncontrolled $o\!\left(2^{-2r}\right)$ term this argument does not apply to the low rate regime, and does, strictly speaking, not rigorously prove concavity at high but finite rates either. It also does not apply to the optimal sensor rules at any finite rate.

In order to make use of Theorem \ref{th:main1} for finite rates, we will instead provide simplified expressions that allow us to prove the concavity of $\mathcal{B}_r^\star$, without explicitly obtaining $\mathcal{B}_r^\star$. This will be accomplished by an argument that is very similar to the argument of Chamberland and Veeravalli \cite{Cham03} [cf.\ \eqref{eq:CHresult}], but applicable to higher rate sensors. Namely, that under some observation models each additional bit allocated to a sensor $S_n$ allows it to close more than half the gap between its current finite rate Bhattacharyya distance and the Bhattacharyya distance of the unquantized observations. The implication of this result is captured by the following lemma.

\begin{lemma}\label{lem:Binf}
If
\begin{equation}
\mathcal{B}^\star_{r}+\mathcal{B}_\infty\leq 2\mathcal{B}^\star_{r+1}
\label{eq:lemBinfStat}
\end{equation}
for all rates $r\in \mathbb{N}$, then $\mathcal{B}^\star_r$ is a discrete concave function of the rate $r$.
\end{lemma}
\begin{proofs}
Since $\mathcal{B}^\star_r$ is a non-decreasing function of the rate $r$ we have
\begin{equation*}\begin{split}
\mathcal{B}^\star_{r}+\mathcal{B}^\star_{r+2}&\leq \mathcal{B}^\star_{r}+\mathcal{B}_\infty \\
&\leq 2\mathcal{B}^\star_{r+1}\,,
\end{split}\end{equation*}
which is the definition of a discrete concave function [cf. \eqref{eq:concdef}]. \qed
\end{proofs}

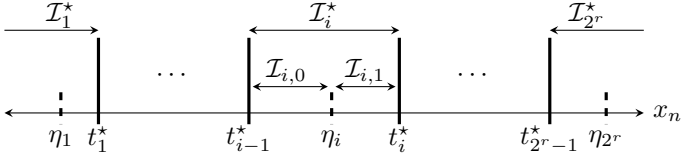
\begin{figure}[t]
\centering
\begin{tikzpicture}[align=center,scale=0.5,>=stealth] 
\draw [<->] (-8.5,0) -- (8.5,0); \node [right] at (8.5,0) {$x_n$};
\draw [very thick,-](-6, 2) -- (-6, -0.3); \node at (-6,-.7) {$t^\star_1$};
\node at (-4,1) {$\dotsb$};
\draw [very thick,-](-2, 2) -- (-2, -0.3);\node at (-2,-0.7){$t^\star_{i-1}$};
\draw [very thick,-](2, 2) -- (2, -0.3);\node at (2,-0.7){$t^\star_{i}$};
\node at (4,1) {$\dotsb$};
\draw [very thick,-](6, 2) -- (6, -0.3);\node at (6,-0.7){$t^\star_{2^r-1}$};
\draw [very thick,dashed,-](0.2, 0.55) -- (0.2, -0.3);\node at (0.2,-0.7){$\eta_{i}$};
\draw [very thick,dashed,-](7.5, 0.55) -- (7.5, -0.3);\node at (7.5,-0.7){$\eta_{2^r}$};
\draw [very thick,dashed,-](-7, 0.55) -- (-7, -0.3);\node at (-7,-0.7){$\eta_{1}$};
\draw [<->] (-2,2.2) -- (2,2.2);\node at (0,2.6) {$\mathcal{I}^\star_i$};
\draw [->] (-8.5,2.2) -- (-6,2.2);\node at (-7,2.6) {$\mathcal{I}^\star_1$};
\draw [<-] (6,2.2) -- (8.5,2.2);\node at (7,2.6) {$\mathcal{I}^\star_{2^r}$};
\draw [<->] (-1.9,0.7) -- (0.1,0.7);\node at (-1,1.1) {$\mathcal{I}_{i,0}$};
\draw [<->] (0.3,0.7) -- (1.9,0.7);\node at (1.1,1.1) {$\mathcal{I}_{i,1}$};
\end{tikzpicture}
\caption{The setting of the thresholds in optimum $r$ bit quantizer and the resulting sub-optimum $r+1$ bit quantizer.}
\label{fig:bindivision}\end{figure}

For many observation distributions $\mathcal{B}_\infty$ admits a closed form expression. In the case of the Laplacian model it holds that $\mathcal{B}_\infty = m - \log(1+m)$ and in case of the Gaussian model it holds that $\mathcal{B}_\infty = m^2/2$, which can be seen from \eqref{eq:BenBhattLapla} and \eqref{eq:BenBhattGauss} by letting $r \rightarrow \infty$. Lemma \ref{lem:Binf} thus provides a simplification towards proving the concavity of $\mathcal{B}^\star_r$. 
Now, for any rate $r$, let the \emph{Bhattacharyya coefficient} corresponding to an optimum Bhattacharyya distance be defined as
\begin{equation*}
b^\star_r\triangleq e^{-\mathcal{B}^\star_r}\,,
\label{eq:bstar}\end{equation*}
and the Bhattacharyya coefficient of an observation be defined as
\begin{equation}
b_\infty\triangleq e^{-\mathcal{B}_\infty}\,.
\label{eq:binfty}\end{equation}
Then
\begin{equation}
b_{r}^\star\,b_\infty\geq \left(b_{r+1}^\star\right)^2
\label{eq:binequality}\end{equation}
is equivalent to \eqref{eq:lemBinfStat} in Lemma~\ref{lem:Binf}. In what follows, now focusing on the Bhattacharyya coefficients, we will propose a sequence of increasingly simplified sufficient conditions under which \eqref{eq:binequality} holds. To this end, let for a given observation model and rate $r$ a given sensor $S$ be an optimally designed monotone quantizer with threshold vector $\underline{t}_r^\star \triangleq \left(t^\star_1,\ldots,t^\star_{2^r-1} \right)$ and intervals $\mathcal{I}^\star_1\triangleq (-\infty,t^\star_1],\ldots,\mathcal{I}^\star_i\triangleq [t^\star_{i-1},t^\star_i],\ldots,\mathcal{I}^\star_{2^r}\triangleq [t^\star_{2^r-1},\infty)$,  which lead to the optimal (minimal) Bhattacharyya coefficient $b^\star_{r}$, or equivalently an optimal (maximal) Bhattacharyya distance $\mathcal{B}^\star_r$, given by
\begin{equation}
b^\star_r=\sum_{i=1}^{2^r} \sqrt{p_0(i)p_1(i)}\,,
\label{eq:bstardef}\end{equation}
where \[p_h(i)\triangleq \Pr\left(X \in \mathcal{I}^\star_i  \vert H=h\right)=\int_{\mathcal{I}^\star_i}\!f_{X \vert H}(x\vert h)\,dx\,,\] for $h\in\{0,1\}$.
Each interval $\mathcal{I}^\star_i=[t^\star_{i-1},t^\star_i]$ can be divided into two sub-intervals $\mathcal{I}_{i,0}\triangleq[t^\star_{i-1},\eta_i]$ and $\mathcal{I}_{i,1}\triangleq[\eta_i,t^\star_i]$, where $t^\star_{i-1}\leq \eta_i\leq t^\star_i$, in order to construct a (not necessarily optimal) rate $r+1$ monotone quantizer, with a Bhattacharyya coefficient $b_{r+1}$ given by
\begin{equation}
b_{r+1}=\sum_{i=1}^{2^r}\sum_{j=0}^{1}\sqrt{p_0(i,j)p_1(i,j)}\,,
\label{eq:bplusdef}\end{equation}
where 
\begin{equation}
p_h(i,j)\triangleq \Pr\left(X\in \mathcal{I}_{i,j} \vert H=h\right)=\int_{\mathcal{I}_{i,j}}\!f_{X\vert H}(x\vert h)\,dx\,,
\label{eq:p_hij}
\end{equation}
for $j,h\in\{0,1\}$. Fig.~\ref{fig:bindivision} illustrates the procedure of creating an $r+1$ bit quantizer from the given optimal $r$ bit quantizer. Now, if for any optimal monotone quantizer of rate $r$ with threshold vector $\underline{t}_r^\star$ and Bhattacharyya coefficient $b^\star_r$, one can build a rate $r+1$ monotone quantizer, as described above, with threshold vector \[\underline{t}_{r+1} \triangleq \left(\eta_1,t^\star_1,\ldots,\eta_i,t^\star_i,\ldots,\eta_{2^r-1},t^\star_{2^r-1},\eta_{2^r} \right)\]
and Bhattacharyya coefficient $b_{r+1}$ that satisfies
\begin{equation}
b_{r}^\star\,b_\infty\geq b_{r+1}^2 \,,
\label{eq:bplus}\end{equation}
then the inequality in \eqref{eq:binequality} will hold as $b_{r+1} > b_{r+1}^\star$, and it will follow that $\mathcal{B}^\star_r$ is a discrete concave function of $r$. 

Verifying \eqref{eq:bplus} through the direct use of \eqref{eq:bstardef} and \eqref{eq:bplusdef} for any threshold vector $\underline{t}_{r+1}$  is a formidable task. However, it turns out that it is sufficient to verify an analogue of \eqref{eq:bplus} for  each possible quantization interval separately. This idea is precisely captured by the following theorem which constitute the main contribution of this section.

\begin{theorem}
Consider a binary hypothesis testing problem with conditional observation distributions $f_{X\vert H}(x\vert h)$, where $x \in \mathbb{R}$ and $h\in\{0,1\}$. Let $\mathcal{I}_i^\star = [t^\star_{i-1},t^\star_i]$ for $i \in \{1,\ldots,2^r\}$ denote the intervals of an optimal monotone quantizer of rate $r$, and let $\mathcal{I}_{i,0}\triangleq [t^\star_{i-1},\eta_i]$ and $\mathcal{I}_{i,1}\triangleq [\eta_i,t^\star_i]$ denote sub-intervals of $\mathcal{I}_i^\star$ obtained for some $\eta_i$. Let $f_h(x|i) \triangleq f_{X|H}(x|h) / p_h(i)$ for $x \in \mathcal{I}_i^\star$ be the density of the observation $X$, conditioned on the true hypothesis $H$ and the event that 
$X \in \mathcal{I}_i^\star$, and let $p_h(j|i) \triangleq \Pr(X\in \mathcal{I}_{i,j}\vert X\in \mathcal{I}^\star_i,H=h)$. Then, the optimum Bhattacharyya distance $\mathcal{B}^\star_r$ is a discrete concave function of the rate $r$ if for each $r \in \mathbb{N}$ and $i \in \{1,\ldots,2^r\}$ there exists an $\eta_i$ for which
\begin{equation}\begin{split}
\left[\sqrt{p_0(0\vert i)p_1(0\vert i)}+\sqrt{p_0(1\vert i)p_1(1\vert i)} \right]^2\leq \hspace{5em}\\
\int_{\mathcal{I}^\star_i}\!\sqrt{f_0\left(x\vert i\right)f_1\left(x\vert i\right)}\,{d}x\, .
\label{eq:th2}\end{split}\end{equation}
\label{th:bstar}\end{theorem}

\begin{remark} It is, as noted before, in general intractable to explicitly find the optimal quantization intervals $\mathcal{I}^\star_i$ for a rate-$r$ sensor. It should however be stressed that this is not required for verification of the conditions of Theorem~\ref{th:bstar}. In fact, if \eqref{eq:th2} holds when $\mathcal{I}^\star_i$ is replaced by an arbitrary interval on the real line, then it holds for $\mathcal{I}^\star_i$ for any $r \in \mathbb{N}$ and $i \in \{1,\ldots,2^r\}$. The value of restricting Theorem \ref{th:bstar} to optimal quantization intervals $\mathcal{I}^\star_i$ is that it allows us to exclude intervals that are for some reason a-priori known to be suboptimal. This observation is used in Section \ref{sec:lapla}, when considering the Laplacian observation model.
\end{remark}

\begin{proofs}See Appendix \ref{app:th3}.
\end{proofs}

\begin{figure*}[!t]
\centering
\begin{minipage}[b]{0.3\linewidth}
\includegraphics[width=\columnwidth]{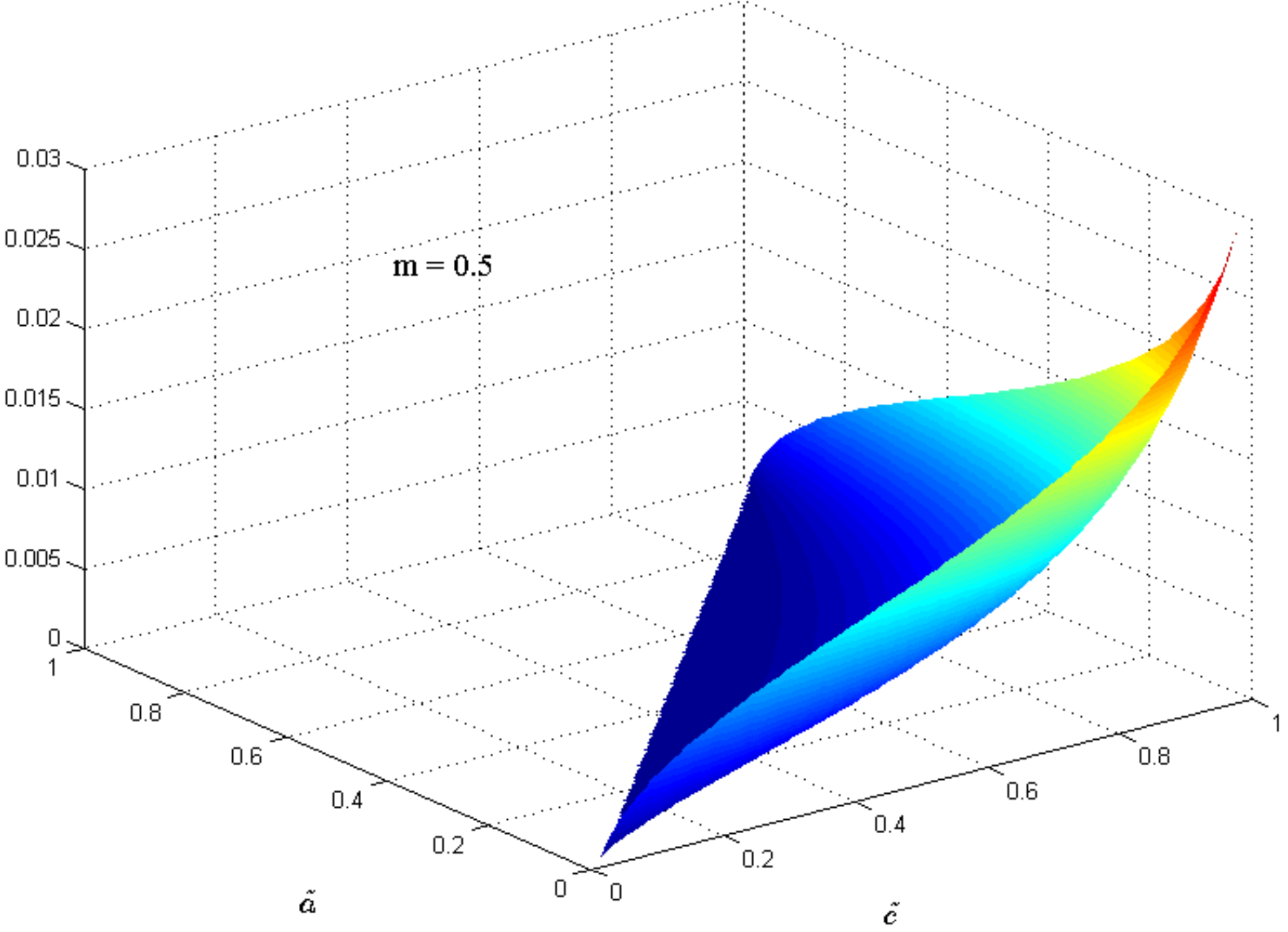}
\end{minipage}
\hspace{0.05cm}
\begin{minipage}[b]{0.3\linewidth}
\includegraphics[width=\columnwidth]{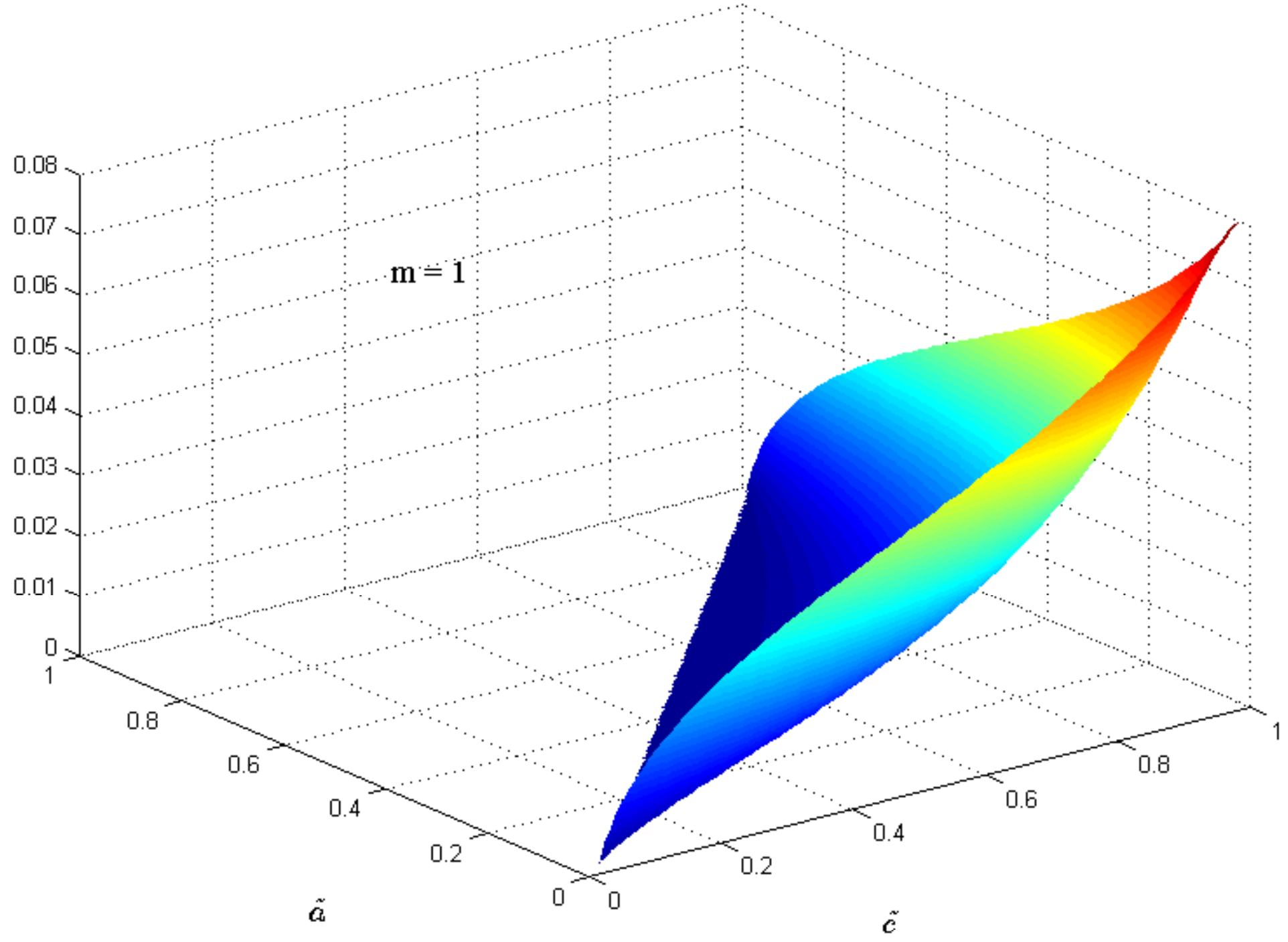}
\end{minipage}
\hspace{0.05cm}
\begin{minipage}[b]{0.3\linewidth}
\includegraphics[width=\columnwidth]{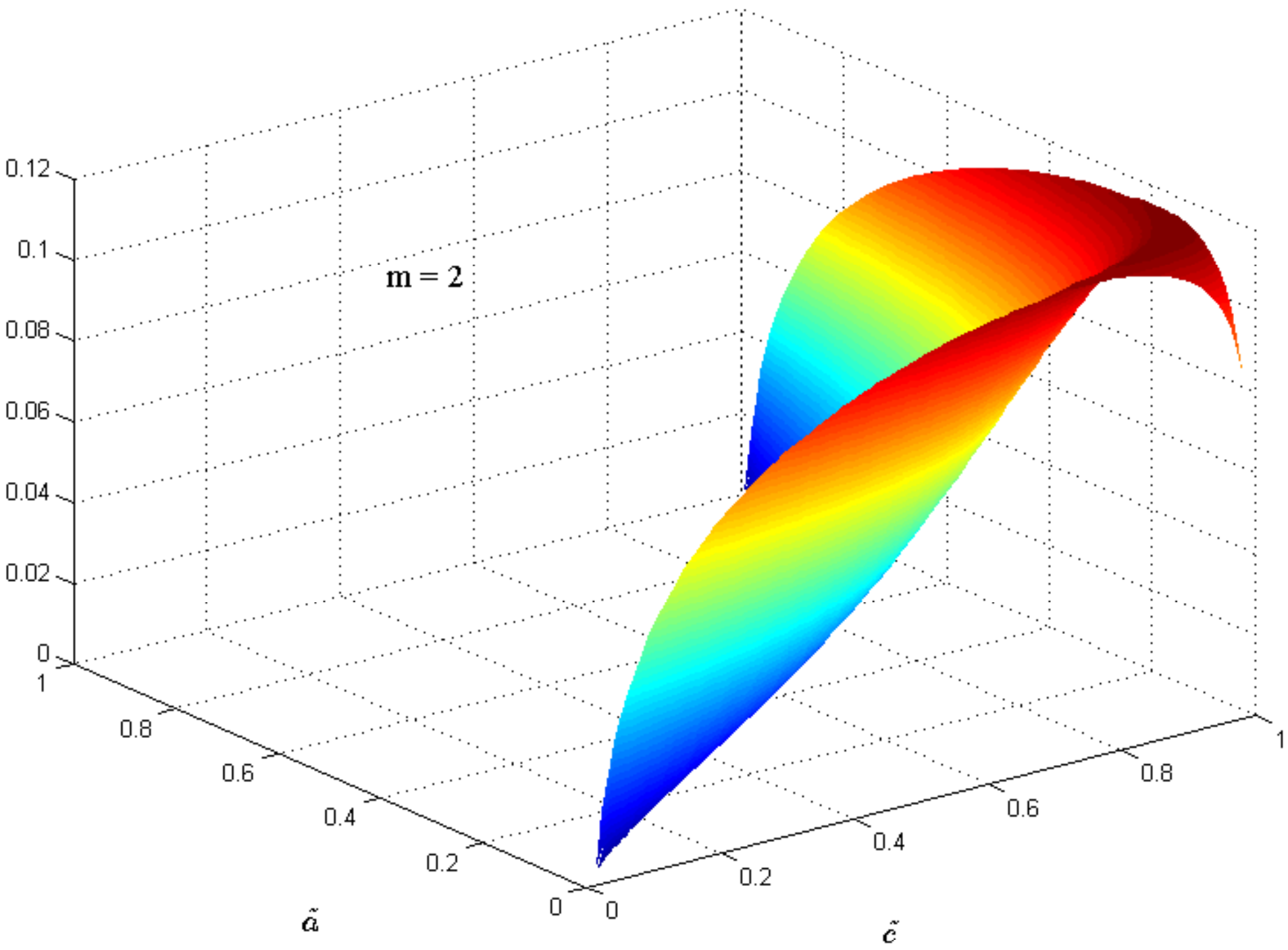}
\end{minipage}
\caption{$b_{\infty\vert i}-\left(b_{1\vert i} \right)^2$ for different intervals $\left[\tilde{a},\tilde{c}\right]$ and for $m=0.5$, $1$ and $2$.}
\label{fig:deltab}
\end{figure*}

In order to shed further light on the conditions posed by Theorem \ref{th:bstar}, let us label the square root of the left-hand-side of the inequality in \eqref{eq:th2} according to
\begin{equation}
b_{1\vert i}\triangleq\sqrt{p_0\!\left(0\vert i\right)p_1\!\left(0\vert i\right)}+\sqrt{p_0\!\left(1\vert i\right)p_1\!\left(1\vert i\right)}\,,
\label{eq:b1idef}\end{equation}
and the right-hand-side of \eqref{eq:th2} according to
\begin{equation}
b_{\infty\vert i}\triangleq\int_{\mathcal{I}^\star_i}\!\sqrt{f_0\left(x\vert i\right)f_1\left(x\vert i\right)}\,{d}x\,.
\label{eq:binfidef}\end{equation}
Comparing with the definition of the Bhattacharyya coefficient of a monotone $r$ bit quantizer, $b_{1\vert i}$ has the following interpretation: It is the Bhattacharyya coefficient of a monotone one bit quantization of $X$, conditioned on $X\in \mathcal{I}^\star_i$. The quantity in \eqref{eq:binfidef}, $b_{\infty\vert i}$, has an analogous interpretation: It is the Bhattacharyya coefficient of the unquantized observation $X$, conditioned on $X\in\mathcal{I}^\star_i$. Expressed in these quantities, the condition in \eqref{eq:th2} can be stated as
\begin{equation*}
b_{1\vert i}^2 \leq b_{\infty\vert i}\,,
\end{equation*}
or equivalently
\begin{equation*}
2\mathcal{B}_{1\vert i} \geq \mathcal{B}_{\infty\vert i}\,,
\end{equation*}
where $\mathcal{B}_{1\vert i} \triangleq -\log b_{1\vert i}$ and $\mathcal{B}_{\infty\vert i} \triangleq  -\log b_{\infty\vert i}$. In words, Theorem \ref{th:bstar} thus states the following: Conditioned on an observation $X$ being in any given interval of the real line (or an interval of an optimum monotone quantizer), if there is a one bit quantization of $X$ with Bhattacharyya distance more than half of the Bhattacharyya distance of $X$ itself, then having rate balanced sensors is optimal. This is in agreement with Chamberland and Veeravalli's result [cf. \eqref{eq:CHresult}]. The conditioning on $X \in \mathcal{I}_i^\star$ is in part what generalizes the result to higher rates. However, verifying this condition is considerably harder than verifying the condition of \cite{Cham03} as it needs to be established for all possible optimal intervals $\mathcal{I}_i^\star$. Nevertheless, we proceed to discuss a few cases where the conditions of Theorem \ref{th:bstar} can be established in practice.

\section{Examples}\label{sec:applic}
In this section, we consider the Laplacian and the Gaussian observation models introduced in \eqref{eq:Laplacase} and \eqref{eq:Gausscase}, respectively. We will first consider the Laplacian case and prove that the inequality in \eqref{eq:th2} is satisfied for any optimal interval of any rate-$r$ sensor. This allows us to use Theorem~\ref{th:bstar} to draw the conclusion that rate balancing is an optimal strategy when the observation model at the sensors is given by \eqref{eq:Laplacase}. We will later conjecture that the same is true for the Gaussian case, and provide our support for this conjecture.
\subsection{Laplacian Observations}\label{sec:lapla}

Before proving the main statement for the Laplacian case, we will briefly discuss an important property of an optimum rate-$r$ monotone quantizer. We will use this property to prove our main statement later in this section. To this end, consider the case where the observation $X$ at sensor $S$ is distributed as in \eqref{eq:Laplacase} with $s=1$, and note again that the assumption of $s=1$ can be made without loss of generality. The likelihood ratio for this observation model is given by
\begin{equation*}
l(x)=\left\{ \begin{array}{ll}
        e^{-2m} &\quad\quad\quad x< -m\,,\\
        e^{2x} & -m \leq x\leq m\,,\\
	  e^{2m} &\quad\quad\quad x> m\, .\end{array} \right.
\end{equation*}
Since the likelihood ratio is constant for $\vert x \vert > m$, no partitioning is needed for $\vert x \vert > m$ \cite{Tsi93, Ben89}. In other words, all the thresholds of an optimum quantizer with rate $r \geq 1$ are in the interval $[-m,m]$, i.e., $-m \leq t^\star_1\leq t^\star_{2^r-1}\leq m$.
\begin{lemma}\label{lem:Laplacian}
Suppose that the observation model at a sensor $S$ is given by \eqref{eq:Laplacase} (with $s=1$). Let $\mathcal{I}^\star_i$ for some $i \in \{1,\ldots,2^r\}$ denote an interval from an optimum monotone quantizer of rate $r$. Then for any interval $\mathcal{I}^\star_i$, there is a threshold $\eta_i\in \mathcal{I}^\star_i$ which divides $\mathcal{I}^\star_i$ into two intervals $\mathcal{I}_{i,0}$ and $\mathcal{I}_{i,1}$ satisfying \eqref{eq:th2}.
\end{lemma}

\begin{proofs}See Appendix \ref{app:lemm4}.
\end{proofs}

\begin{figure*}[!t]
\centering
\begin{minipage}[b]{0.48\linewidth}
\includegraphics[width=\columnwidth]{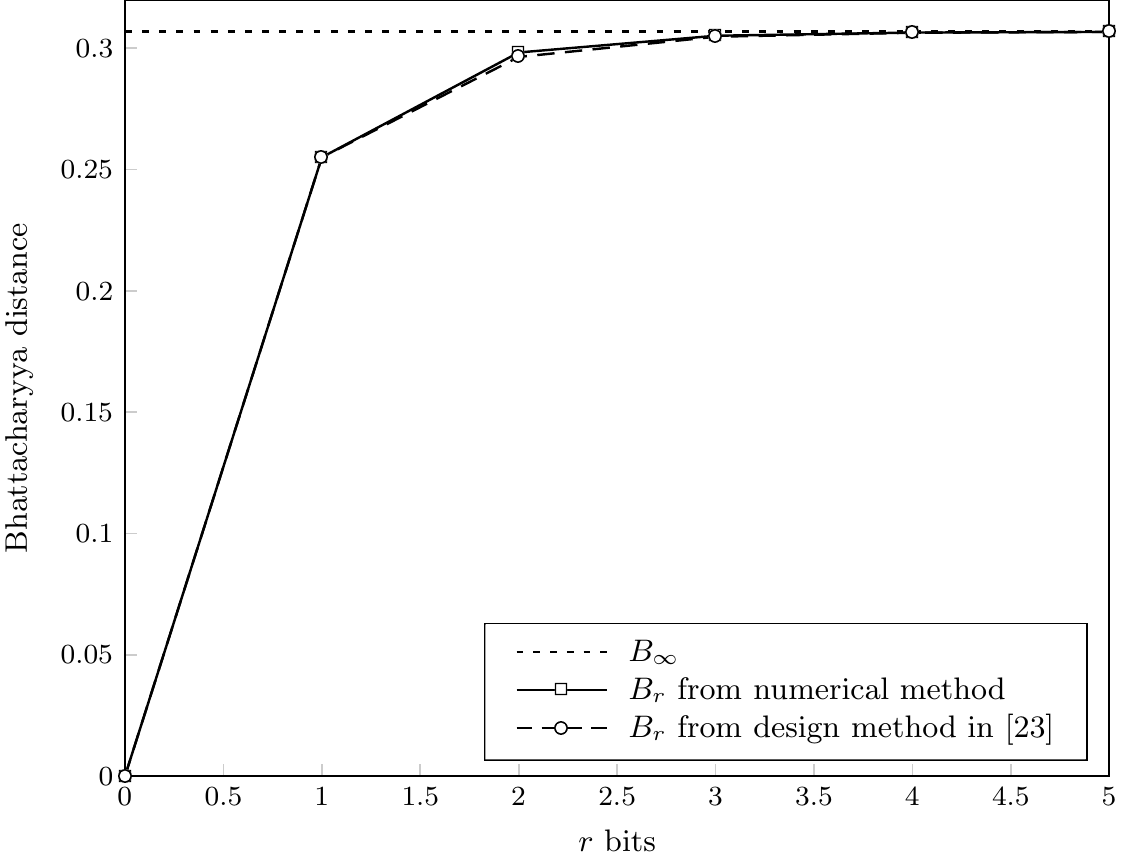}
\end{minipage}
\hspace{0.05cm}
\begin{minipage}[b]{0.48\linewidth}
\includegraphics[width=\columnwidth]{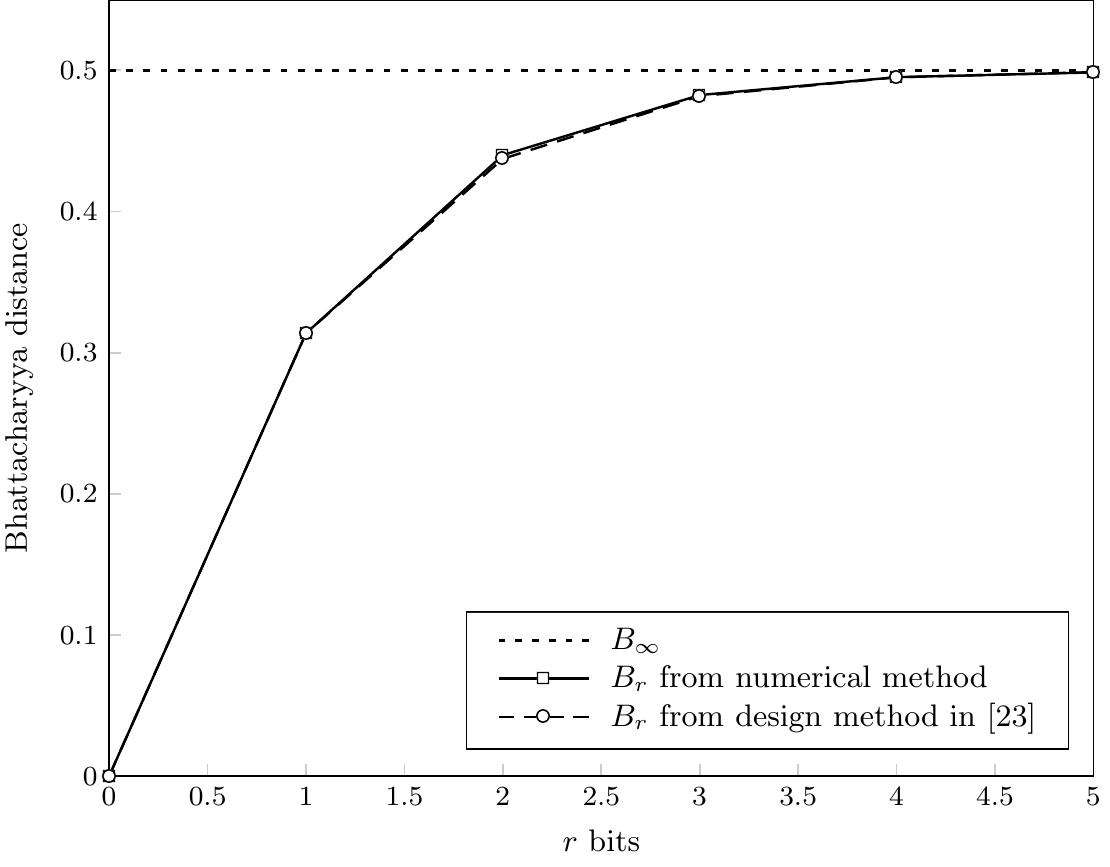}
\end{minipage}
\caption{Bhattacharyya distance of a sensor designed using our numerical method and the method proposed in \cite{Ben89} and the Bhattacharyya distance contained in each observation, for the Laplacian case (left) and the Gaussian case (right), when $m=1$.}
\label{fig:BGaussLapla}
\end{figure*}

\subsection{Gaussian Observations}
When the observations at the sensors are Gaussian distributed as \eqref{eq:Gausscase} it is, similar to the Laplacian case, in principle sufficient to show the condition \eqref{eq:th2} holds for optimum intervals in $\mathcal{X}=\mathbb{R}$. Let $\mathcal{I}^\star_i\triangleq \left[a,c\right]$, where $-\infty \leq a< c\leq \infty$. In this case the LLR is not bounded, which implies that there is no a-priori limitation on the possible intervals as there was in the Laplacian case. Using \eqref{eq:concatenatedprob} we obtain
\begin{equation}\begin{split}
&b_{\infty\vert i}=\frac{e^{-\frac{m^2}{2}}\big[\mathcal{Q}(a)-\mathcal{Q}(c)\big]}{\sqrt{\big[ \mathcal{Q}(a-m)-\mathcal{Q}(c-m)\big]\big[ \mathcal{Q}(a+m)-\mathcal{Q}(c+m)\big]\,}}
\label{eq:Gaussbinf}\end{split}\end{equation}
and
\begin{equation}\begin{split}
b_{1\vert i}&=\\
&\sqrt{\left[\frac{\mathcal{Q}(a-m)-\mathcal{Q}(\eta_i-m)}{\mathcal{Q}(a-m)-\mathcal{Q}(c-m)}\right]\!\left[\frac{\mathcal{Q}(a+m)-\mathcal{Q}(\eta_i+m)}{\mathcal{Q}(a+m)-\mathcal{Q}(c+m)}\right]}\\
+&\sqrt{\left[\frac{\mathcal{Q}(\eta_i-m)-\mathcal{Q}(c-m)}{\mathcal{Q}(a-m)-\mathcal{Q}(c-m)}\right]\!\left[\frac{\mathcal{Q}(\eta_i+m)-\mathcal{Q}(c+m)}{\mathcal{Q}(a+m)-\mathcal{Q}(c+m)}\right]} \, ,
\label{eq:Gaussbi}\end{split}\end{equation}
where $\eta_i$ is a threshold for which $a\leq \eta_i \leq c$, and where $\mathcal{Q}$ denotes the Gaussian tail probability defined in \eqref{eq:Qfun}. It is worth noting that when the rate $r=0$, i.e., $a=-\infty, c=\infty$, the conditional Bhattacharyya coefficients in \eqref{eq:Gaussbinf} and \eqref{eq:Gaussbi} reduce to the Bhattacharyya coefficient of the raw observation given by $b_\infty=\exp\left(-\frac{m^2}{2}\right)$, and for $\eta_1=0$ the Bhattacharyya coefficient in one bit quantized observation is given by $b_1=2\sqrt{\mathcal{Q}(-m)\mathcal{Q}(m)}$. For this particular case, it was already shown in \cite{Cham03} that $\left( b_1\right)^2\leq b_\infty$, which means \eqref{eq:th2} holds for $r=0$ and some $\eta_1$. We \emph{conjecture} that \eqref{eq:th2} is in fact true for all $a,c \in \mathbb{R}$ with $a < c$, $m > 0$, and $r > 0$ and some $\eta_i$, but we have not been able to prove this stringently.

Although $\mathcal{Q}(y)$ is a well defined function, it is hard to further simplify \eqref{eq:Gaussbinf} and \eqref{eq:Gaussbi} for a given $\eta_i$ and our attempt to prove that the inequality \eqref{eq:th2} holds for some $\eta_i$ has been in vain. Therefore, and without any proof in the following, we will introduce a suggested threshold $\eta_i$ for which the inequality \eqref{eq:th2} has been numerically shown to hold over a large range of choices for $a,c$ and $m$. To this end, consider an interval $\mathcal{I}^\star_i\triangleq \left[a,c\right]$ and conditional observation distributions given by [cf. \eqref{eq:concatenatedprob}]
\begin{equation}\begin{split}
f_0\left(x\vert i\right)&=\frac{e^{-\frac{\left(x+m\right)^2}{2}}}{\sqrt{2\pi}\,\big[ \mathcal{Q}(a+m)-\mathcal{Q}(c+m) \big]}\,,\\
f_1\left(x\vert i\right)&=\frac{e^{-\frac{\left(x-m\right)^2}{2}}}{\sqrt{2\pi}\,\big[ \mathcal{Q}(a-m)-\mathcal{Q}(c-m) \big]}\,.
\label{eq:concatf}\end{split}\end{equation}
The equal likelihood ratio threshold $x_\mathrm{L}$ under which \[f_1\left(x_\mathrm{L}\vert i\right)=f_0\left(x_\mathrm{L}\vert i\right)\,,\]
is given by
\begin{equation}
x_\mathrm{L}=\frac{1}{2m}\ln\frac{\mathcal{Q}(a-m)-\mathcal{Q}(c-m)}{\mathcal{Q}(a+m)-\mathcal{Q}(c+m)}\,.
\label{eq:LRth}\end{equation}
Using simulations we have numerically observed that by choosing $\eta_i=x_\mathrm{L}$ for \emph{any} interval $\left[a,c\right]$, and therefore also any optimal intervals, the inequality \eqref{eq:th2} holds. Using the logit transforms
\begin{equation*}
a=\log\frac{\tilde{a}}{1-\tilde{a}} \quad \text{and} \quad
c=\log\frac{\tilde{c}}{1-\tilde{c}}
\end{equation*}
allows us to parameterize $a \in \mathbb{R}$ and  $c \in \mathbb{R}$ with $a < c$ by $0 \leq \tilde{a} < \tilde{c} \leq 1$. Fig.~\ref{fig:deltab} illustrates $b_{\infty\vert i}-b_{1\vert i}^2$ for the cases that $m=0.5,1,2$, and for $0 \leq \tilde{a} < \tilde{c} \leq 1$. It can be seen (numerically) that $b_{\infty\vert i}-b_{1\vert i}^2 \geq 0$, or equivalently that $b_{\infty\vert i} \geq b_{1\vert i}^2$ for all $a \leq c$ or $0 \leq \tilde{a} \leq \tilde{c} \leq 1$.

We have, besides the plots shown herein, also numerically evaluated $b_{\infty\vert i}-b_{1\vert i}^2$ for a much larger range of different values of $m$, and our simulation results consistently support the conjecture that by choosing $\eta_i=x_\mathrm{L}$ the inequality \eqref{eq:th2} always holds. Thus, based on this numerical evidence, we have good reason to believe that rate balancing is an optimal strategy also when the observations at the sensors are modeled as in \eqref{eq:Gausscase}, i.e., under additive Gaussian noise.

It is also worth to note that by following the instructions in \cite{Ben89} using an asymptotically optimal companding function $q \,:\, \mathcal{I}_i^\star \mapsto [0,1]$ we can acquire another threshold for an arbitrary interval $\left[a,c\right]$ as $x_\mathrm{B} = q^{-1}(1/2)$ which can be explicitly expressed as
\[x_\mathrm{B}=\sqrt{3}\,\mathcal{Q}^{-1}\left(\frac{1}{2}{\mathcal{Q}\left(\frac{a}{\sqrt{3}}\right)}+\frac{1}{2}{\mathcal{Q}\left(\frac{c}{\sqrt{3}}\right)} \right)\,.\] Note here that $x_\mathrm{B} \neq x_\mathrm{L}$ in general. Our simulation results indicate that \eqref{eq:th2} also holds for $\eta_i=x_\mathrm{B}$. It remains an open problem to stringently prove \eqref{eq:th2} for $\eta_i = x_\mathrm{L}$ or $\eta_i = x_\mathrm{B}$, or to identify another choice of $\eta_i$ for which \eqref{eq:th2} is more amendable to be proven.

\section{Simulation Results and Discussions}\label{sec:disc}
In this section we present some numerical examples to show the application of our results. We further show how the performance of different rate allocations can be partially compared using \emph{majorization theory} \cite{Mar10} and also the concavity properties of the Bhattacharyya distance.

\begin{figure*}[!t]
\centering
\begin{minipage}[b]{0.48\linewidth}
\includegraphics[width=\columnwidth]{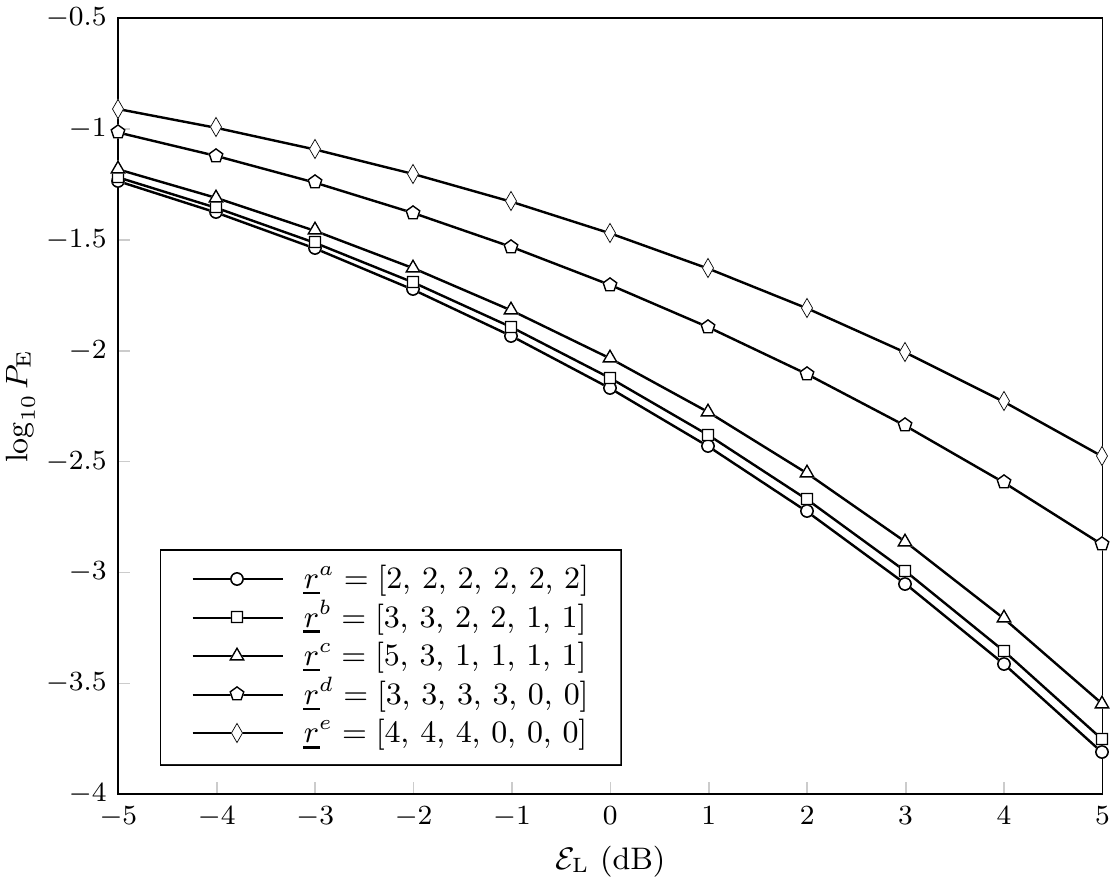}
\caption{Error probability performance of designed sensor networks with different rate allocation schemes and Laplacian observations, as a function of channels SNR $\mathcal{E}_\mathrm{L}$, for $N=6$ sensors and $R=12$ bits per unit time}.
\label{fig:LaplaPE}
\end{minipage}
\hspace{0.2cm}
\begin{minipage}[b]{0.48\linewidth}
\includegraphics[width=\columnwidth]{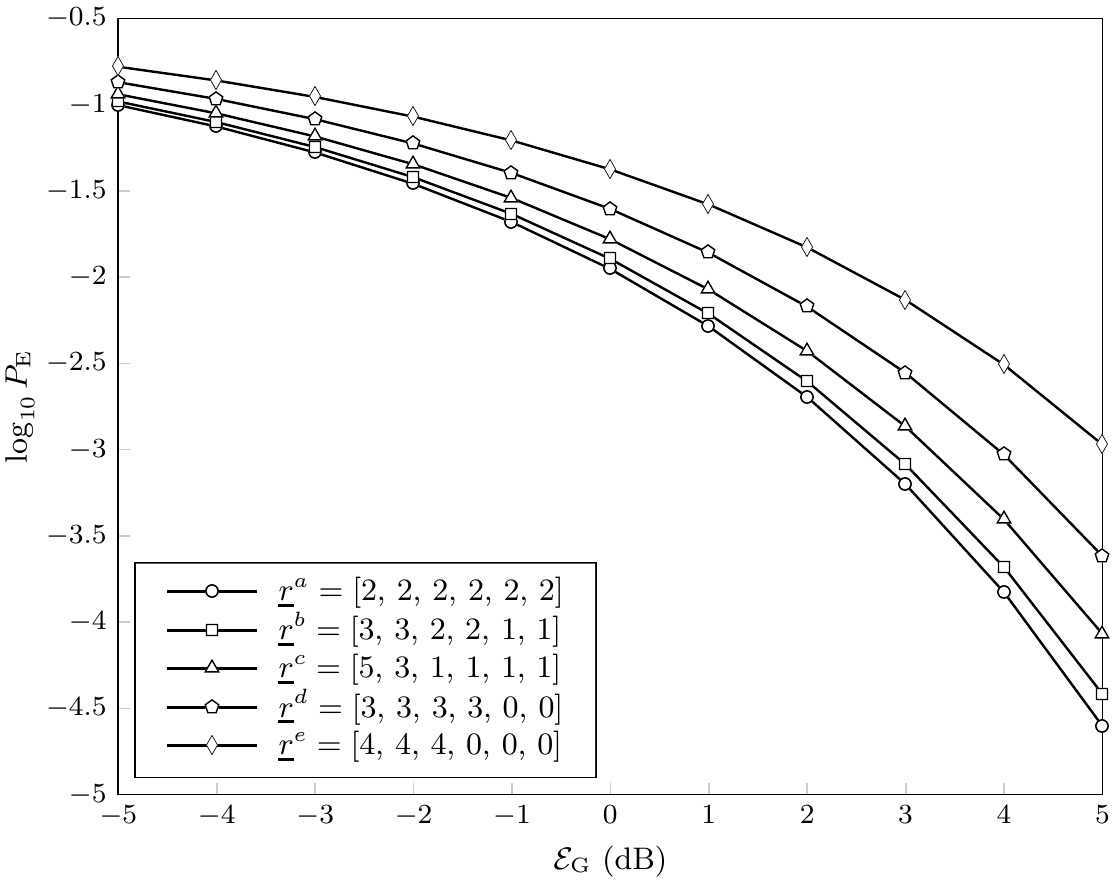}
\caption{Error probability performance of designed sensor networks with different rate allocation schemes and Gaussian observations, as a function of channels SNR $\mathcal{E}_\mathrm{G}$, for $N=6$ sensors and $R=12$ bits per unit time}.
\label{fig:GaussPE}
\end{minipage}
\end{figure*}

Using a numerical method similar to the coordinate descent design method used in \cite{Ald04}, we design quantizers for both the Laplacian and Gaussian observation models in \eqref{eq:Laplacase} and \eqref{eq:Gausscase}, and for different rates, as follows. We first generate $2^r-1$ thresholds $\{t_1,\ldots,t_{2^r-1}\}$ uniformly at random for a monotone quantizer of rate $r$ in an interval $[t_0,t_{2^r}]$, where the interval $[t_0,t_{2^r}]$ is defined symmetrically around zero and contains $0.999$ of the whole probability of $X$, and order the thresholds so that $t_0 \leq t_1\leq\ldots\leq t_{2^r-1}\leq t_{2^r}$. 
Then, in an iterative manner, we modify the position of each threshold -- from $t_1$ to $t_{2^r-1}$ -- while the other thresholds are kept fixed. The position of each threshold, say $t_i$, is modified in the interval $[t_{i-1},t_{i+1}]$ in such a way that the Bhattacharyya distance is numerically maximized. We iteratively modify the position of thresholds until the improvement in the Bhattacharyya distance over a complete pass over all the thresholds is less than $10^{-4}$.

Fig.~\ref{fig:BGaussLapla} illustrates the Bhattacharyya distance of an observation for the Laplacian and Gaussian cases when $m=1$, the maximum Bhattacharyya distance of a designed sensor using the prescribed numerical method, and the Bhattacharyya distance of a designed sensor using the asymptotically optimal method proposed in \cite{Ben89}. The merits of the numerical method is illustrated in this figure, although the difference with respect to the asymptotically optimal design methods is marginal. This figure also illustrates that the numerically obtained Bhattacharyya distances are indeed concave functions, as predicted by our analytical results.

Next, consider a network of $N$ sensors with a given rate allocation $\underline{r}=[r_1, \ldots, r_N]$ that satisfies the rate constraint \eqref{eq:Rconstraint}. Let the maximum Bhattacharyya distance of a single sensor with rate $r_n$ be $\mathcal{B}^\star_{r_n}$ and let the total Bhattacharyya distance of the network with rate allocation $\underline{r}$ be $\mathcal{B}^\star_{\underline{r}}$. The concavity of $\mathcal{B}^\star_{r_n}$ implies that the total Bhattacharyya distance of the network 
\[\mathcal{B}^\star_{\underline{r}}=\sum_{n=1}^N\mathcal{B}^\star_{r_n}\] 
is Schur-concave \cite{Mar10} and consequently, if a rate allocation $\underline{r}^a\triangleq[r^a_1,\ldots,r^a_N]$ is majorized by another rate allocation $\underline{r}^b\triangleq[r^b_1,\ldots,r^b_N]$ -- written as $\underline{r}^a \prec\underline{r}^b$ -- then \[\mathcal{B}^\star_{\underline{r}^a} \geq \mathcal{B}^\star_{\underline{r}^b}\,.\]
The condition $\underline{r}^a\prec \underline{r}^b$ is equivalent to
\begin{equation*}\begin{split}
\sum_{i=1}^n r^a_{[i]}&\leq \sum_{i=1}^n r^b_{[i]}\,, \quad n=1,\ldots,N-1\, , \\
\sum_{i=1}^N r^a_{[i]}&= \sum_{i=1}^N r^b_{[i]}\,,
\end{split}\end{equation*}
where $r^a_{[1]}\geq \ldots \geq r^a_{[N]}$ and $r^b_{[1]}\geq \ldots \geq r^b_{[N]}$. This result is in line with the Theorem \ref{th:main1}: The rate allocation of a rate balanced network is majorized by any other rate allocation and so it follows that its total Bhattacharyya distance is more than (or at least equal to) that of any other network with the same size $N$ and the same rate constraint $R$. Further, majorization provides a tool to compare the performance of different networks by considering the majority of their rate allocation vectors. In what follows, we exemplify these results, again in the Laplacian and the Gaussian cases in two different setups.

\begin{figure*}[!t]
\centering
\begin{minipage}[b]{0.48\linewidth}
\includegraphics[width=\columnwidth]{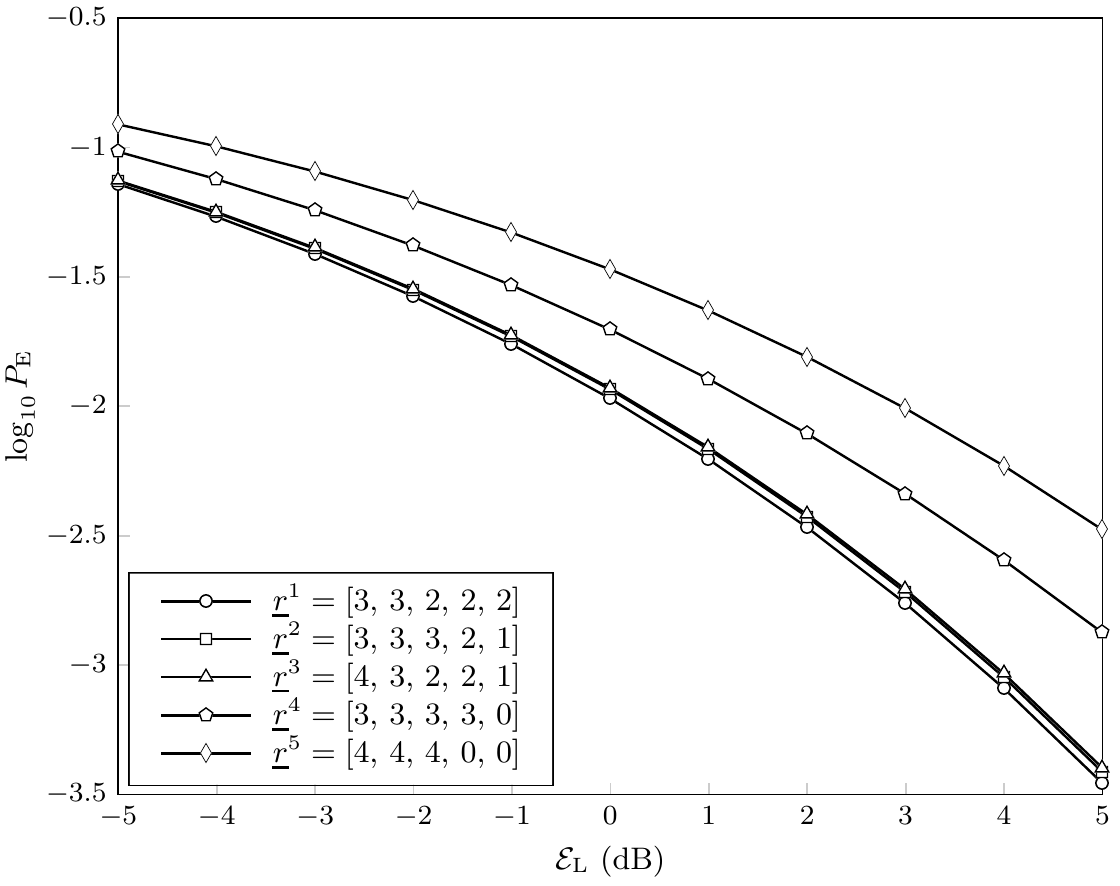}
\caption{Error probability performance of designed sensor networks with different rate allocation schemes and Laplacian observations, as a function of channels SNR $\mathcal{E}_\mathrm{L}$, for $N=5$ sensors and $R=12$ bits per unit time}.
\label{fig:LaplaPE2}
\end{minipage}
\hspace{0.2cm}
\begin{minipage}[b]{0.48\linewidth}
\includegraphics[width=\columnwidth]{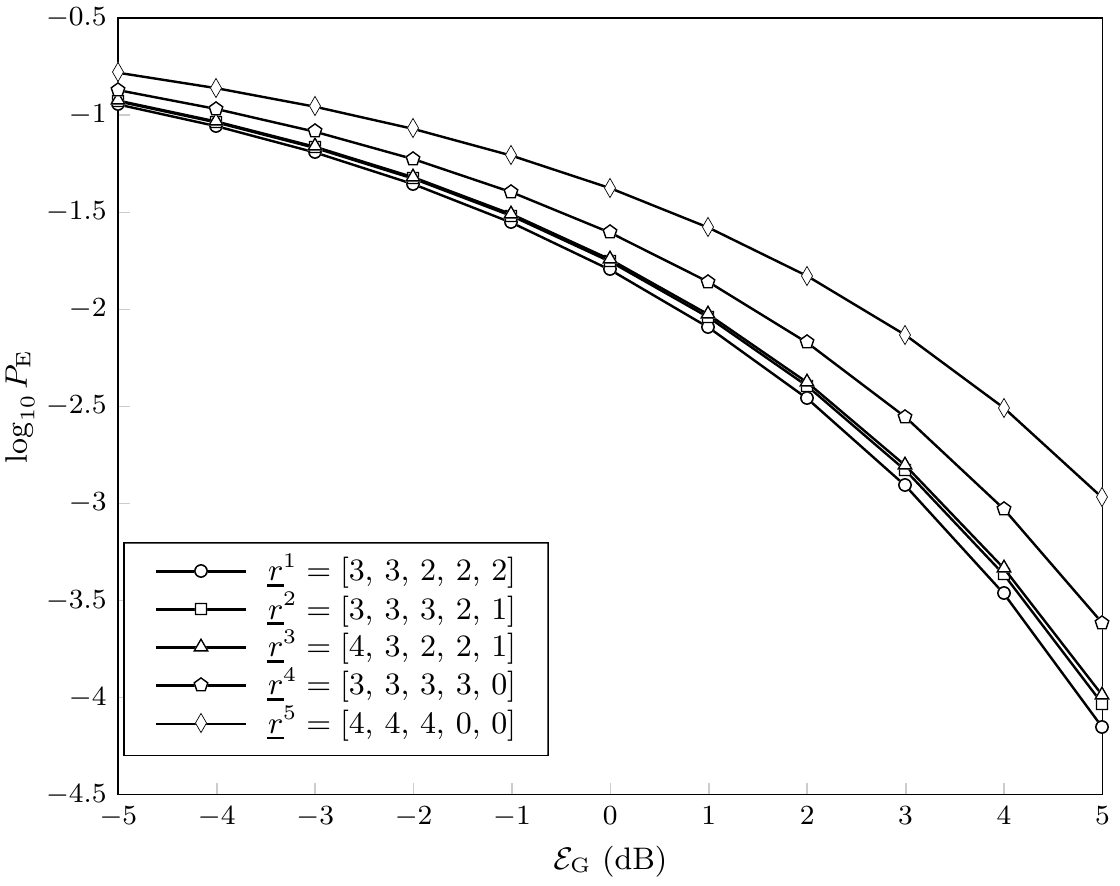}
\caption{Error probability performance of designed sensor networks with different rate allocation schemes and Gaussian observations, as a function of channels SNR $\mathcal{E}_\mathrm{G}$, for $N=5$ sensors and $R=12$ bits per unit time}.
\label{fig:GaussPE2}
\end{minipage}
\end{figure*}

In the first setup, consider a network with a maximum number of sensors $N=6$ arranged as in Fig.~\ref{fig:topology}, where the MAC channel is capable of carrying $R=12$ bits per unit time. We will consider five different rate allocation schemes which all satisfy the rate constraint \eqref{eq:Rconstraint} with rate allocations as follows:
\begin{equation*}\begin{split}
\underline{r}^a&=[2,2,2,2,2,2]\,,\\
\underline{r}^b&=[3,3,2,2,1,1]\,,\\
\underline{r}^c&=[5,3,1,1,1,1]\,,\\
\underline{r}^d&=[3,3,3,3,0,0]\,,\\
\underline{r}^e&=[4,4,4,0,0,0]\,.
\end{split}\end{equation*}
Using the numerical design method described above we designed the sensors for each network and compared their performance in terms of the error probability at the FC under the assumption of equally likely hypotheses, i.e., $\pi_0 = \pi = \tfrac{1}{2}$. In spite of the fact that the Bhattacharyya distance only provides an upper bound on the error probability according to $P_{\mathrm{E}} \leq \sqrt{\pi_0 \pi_1} e^{-\mathcal{B}_{\underline{r}}^\star}$, we see that the results predicted by the analysis of $\mathcal{B}_{\underline{r}}^\star$ hold true also for $P_{\mathrm{E}}$. It is worthwhile to note that we for the purpose of the numerical example can obtain the probability of error $P_\mathrm{E}$ at the FC according to (for more details, see \cite{Alla14})
\begin{equation*} 
P_\mathrm{E}=1-\frac{1}{2}\sum_{\underline{u}} \max_{h=0,1}\left\{P_{\underline{U}\vert H}\left( \underline{u}\vert h\right)\right\},
\end{equation*}
for different per channel signal-to-noise ratios (SNR), without the need for Monte Carlo simulations. Fig.~\ref{fig:LaplaPE} and Fig.~\ref{fig:GaussPE} illustrate the error probability performance of different rate allocation schemes, as a function of the per channel SNR, where the SNR for the Laplacian case is given by $\mathcal{E}_\mathrm{L}\triangleq \vert m \vert ^2 /2$, and for the Gaussian case by $\mathcal{E}_\mathrm{G}\triangleq \vert m \vert ^2$.

Rate balanced sensors (with rate allocation $\underline{r}^a$ which is majorized by the other rate allocations) yields the best performance in all cases. Rate allocation $\underline{r}^b$, which is majorized by rate allocations $\underline{r}^c$, $\underline{r}^d$ and $\underline{r}^e$, has the second best performance among different introduced schemes. In the same way, $\underline{r}^d$ is majorized by $\underline{r}^e$ and has better performance. Since majorization only provides a partial ordering, it is not in general feasible to compare the performance of every pair of rate allocations. For instance, the performances of $\underline{r}^c$ and $\underline{r}^d$ are not comparable using majorization since neither $\underline{r}^c\prec \underline{r}^d$ nor $\underline{r}^d\prec \underline{r}^c$ holds. However, using the concavity properties of the Bhattacharyya distance one can still compare their performance. To this end, consider two networks with rate allocations $\underline{r}^c$ and $\underline{r}^d$, and their corresponding total Bhattacharyya distances $\mathcal{B}^\star_{\underline{r}^c}$ and $\mathcal{B}^\star_{\underline{r}^d}$. By the fact that having two rate-one sensors is better than having one rate-three sensor, i.e., $\mathcal{B}^\star_{(1,1)}\geq\mathcal{B}^\star_{3}$, we obtain
\begin{equation*}\begin{split}
\mathcal{B}^\star_{\underline{r}^c}&=\mathcal{B}^\star_{(1,1)}+\mathcal{B}^\star_{(1,1)}+\mathcal{B}^\star_{3}+\mathcal{B}^\star_{5}\\
&\geq \mathcal{B}^\star_{3}+\mathcal{B}^\star_{3}+\mathcal{B}^\star_{3}+\mathcal{B}^\star_{3}=\mathcal{B}^\star_{\underline{r}^d}\,,
\end{split}\end{equation*}
where $\mathcal{B}^\star_{(1,1)}= \mathcal{B}^\star_{1}+\mathcal{B}^\star_{1}$. 

In the previous example, we have considered the case where the number of sensors $N$ divides the total sum rate of the MAC channel $R$, i.e., $R=kN$. In that case a uniform rate allocation $\underline{r}=[k,k,\ldots,k]$ which is majorized by any other rate allocation has been shown to have the best error probability performance for both observation models \eqref{eq:Laplacase} and \eqref{eq:Gausscase}. Now consider a case where the total number of sensors does not divide the sum rate capacity. In this case an optimal strategy is a non-uniform rate allocation that satisfies condition \eqref{eq:ratebalance}, known as a balanced rate allocation scheme. Let consider the case where $N=5$ sensors arranged as in Fig.~\ref{fig:topology} and the MAC channel is capable of carrying $R=12$ bits per unit time. Consider the following rate allocation schemes which satisfy the rate constraint condition \eqref{eq:Rconstraint}.
\begin{equation*}\begin{split}
\underline{r}^1&=[3,3,2,2,2]\,,\\
\underline{r}^2&=[3,3,3,2,1]\,,\\
\underline{r}^3&=[4,3,2,2,1]\,,\\
\underline{r}^4&=[3,3,3,3,0]\,,\\
\underline{r}^5&=[4,4,4,0,0]\,.
\end{split}\end{equation*}
In this setup, we also observe from Fig.~\ref{fig:LaplaPE2} and Fig.~\ref{fig:GaussPE2} that $\underline{r}^1$ which is majorized by the other rate allocations has the best error probability performance. In fact, comparing to \eqref{eq:ratebalance}, $\underline{r}^1$ is the balanced rate allocation scheme which has been proven to have the optimal performance. In the same way as in the previous example, we can compare the performance of different schemes using the majorization theory and the discrete concavity properties of the Bhattacharyya distance. By doing so, we see that $\underline{r}^2$ has better performance than $\underline{r}^3$, and $\underline{r}^3$ has better performance than $\underline{r}^4$, which itself outperforms $\underline{r}^5$. Thus, to conclude, we see that majorization theory and the discrete concavity properties of the Bhattacharyya distance with respect to the sensor rate provides a powerful and practical tool to compare different rate allocations in wireless sensor networks.
\begin{figure*}[!t]
\centering
\begin{minipage}[b]{0.48\linewidth}
\includegraphics[width=\columnwidth]{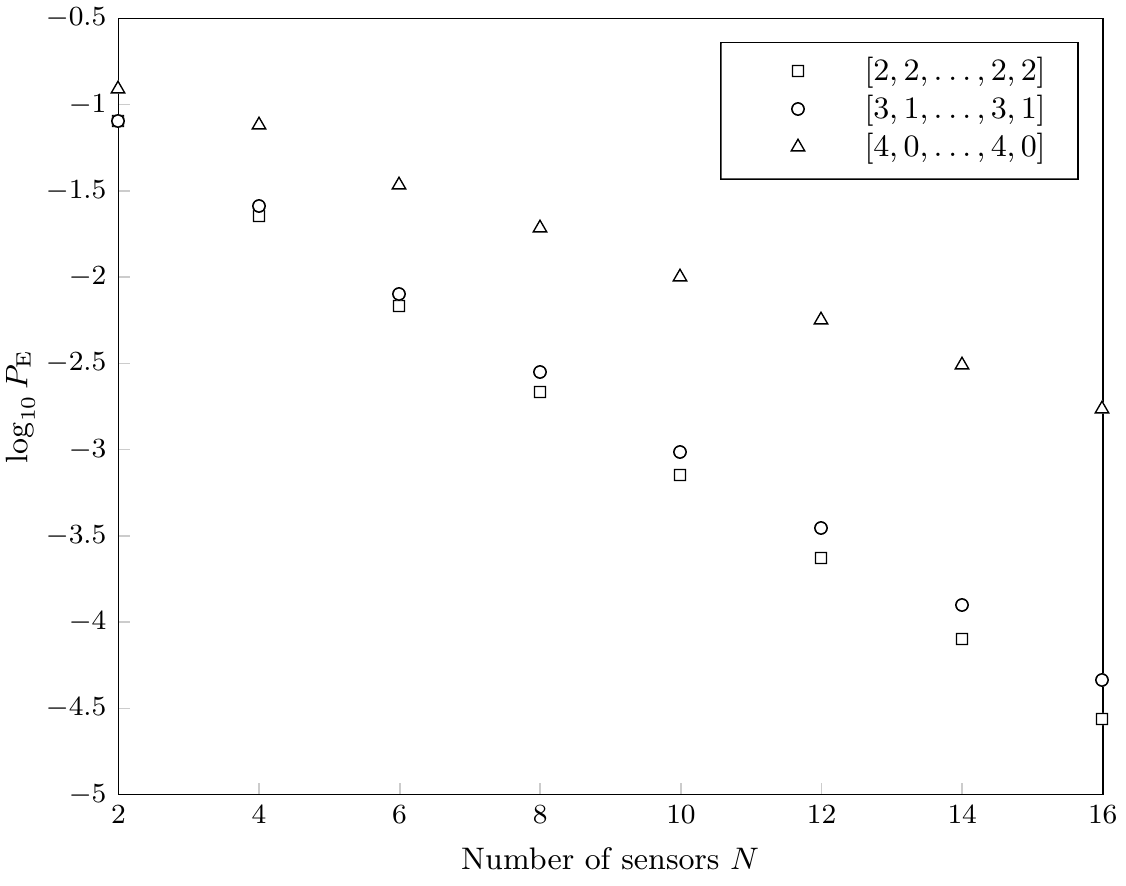}
\caption{Evolution of error probability performance of designed sensor networks of $R=2N$ bits per unit time with different rate allocation schemes and Laplacian observations with SNR $\mathcal{E}_\mathrm{L}=0$ dB, as a function of number of sensors $N$}.
\label{fig:Lapladecay}
\end{minipage}
\hspace{0.2cm}
\begin{minipage}[b]{0.48\linewidth}
\includegraphics[width=\columnwidth]{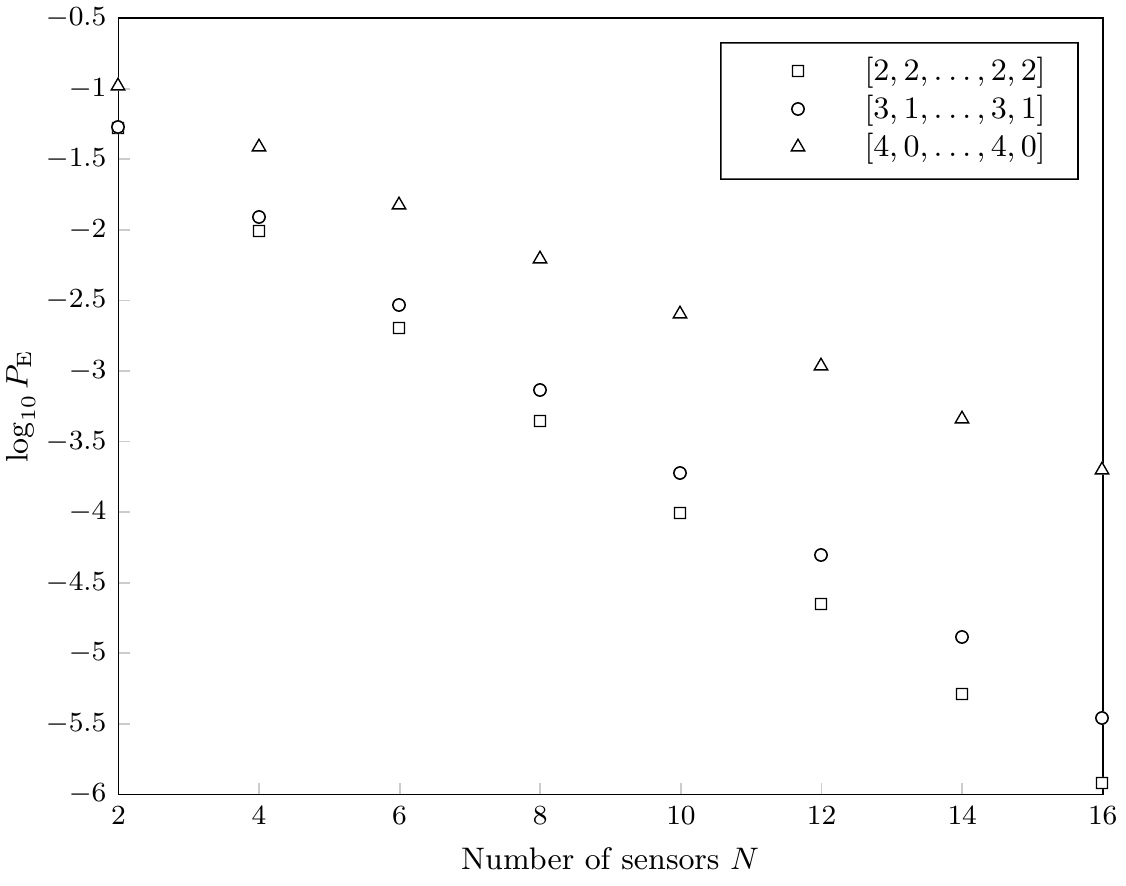}
\caption{Evolution of error probability performance of designed sensor networks of $R=2N$ bits per unit time with different rate allocation schemes and Gaussian observations with SNR $\mathcal{E}_\mathrm{G}=0$ dB, as a function of number of sensors $N$}.
\label{fig:Gaussdecay}
\end{minipage}
\end{figure*}

Now let us assume that $R=2N$ for even $N$ ($N=2L$), and let us consider three different rate allocation schemes, which satisfy the condition in \eqref{eq:Rconstraint}. In the first scheme, we let all the sensors have the same rate $r_n=2$ bits, $n=1,\ldots,N$. In the second scheme, we divide the sensors into two equally-sized groups of three bit sensors and one bit sensors, i.e., $r_{2n-1}=3$ and $r_{2n}=1$ for $n=1,\ldots,L$. In the third scheme, we divide the sensors into two equally-sized groups of four bit sensors and zero bit sensors, i.e., $r_{2n-1}=4$ and $r_{2n}=0$ for $n=1,\ldots,L$. In other words, in the last scheme half of the sensors are turned off. We label three aforementioned schemes respectively as
\[[2,2,\ldots,2,2], \quad [3,1,\ldots,3,1],\quad[4,0,\ldots,4,0]\,.\] In Fig.~\ref{fig:Lapladecay} and Fig.~\ref{fig:Gaussdecay} the evolution of error probability performance of different setups as a function of total number of sensors $N$ is shown. We observe from these figures that the uniform rate allocation scheme not only results in the best error probability performance for any $N$, it also has the best error exponent (decay rate as a function of total number of sensors in a network), as predicted by the superior Bhattacharyya distance.

\section{Conclusion} \label{sec:conc}
In this paper, we considered a decentralized hypothesis testing problem in which a fixed number of sensors send quantized information towards a fusion center through a multiple access channel. We considered the case where the sensors make conditionally independent and identically distributed observations of the true hypothesis, and assumed the MAC channel could be described by a sum rate constraint with total rate $R$. This problem was first considered by Chamberland and Veeravalli under the assumption on an unlimited number of sensors. They provided sufficient conditions under which a Chernoff information optimal strategy is to have $N=R$ rate-one sensors. Since it might not always be feasible to have a large number of sensors in the network, we extended this result to the scenario of an a-priori limited set of sensors and found sufficient conditions under which \emph{rate balancing} is a Bhattacharyya distance optimal strategy. The sufficient conditions were then conclusively proven under a Laplacian additive noise observation model, and conjectured to hold under the more common Gaussian additive noise model. Overall, these results provide a powerful tool applicable to the design of distributed wireless sensor networks.

\appendices

\section{Proof of Lemma \ref{lem:bhatt}}\label{app:lemm1}
For the Bhattacharyya distance of a network of sensors with independent observations we obtain
\begin{equation*}\begin{split}
\mathcal{B}_{\underline{r}}\left(\,\underline{\gamma}\,\right)
&=-\log\left[ \sum_{\underline{u}\in\, \underline{\mathcal{U}}} \sqrt{P_{\underline{U}\vert H}\left(\underline{u}\vert 0\right)\,P_{\underline{U}\vert H}\left(\underline{u}\vert 1\right)}\,\right ]\\
&=-\log\left[\prod_{n=1}^N \sum_{u_n\in\,\mathcal{U}_{r_n}} \sqrt{P_{U_n\vert H}\left({u_n}\vert 0\right)\,P_{U_n\vert H}\left({u_n}\vert 1\right)}\,\right ]\\
&=\sum_{n=1}^N-\log\left[ \sum_{u_n\in\,\mathcal{U}_{r_n}} \sqrt{P_{U_n\vert H}\left({u_n}\vert 0\right)\,P_{U_n\vert H}\left({u_n}\vert 1\right)}\,\right ]\\
&=\sum_{n=1}^N\mathcal{B}_{r_n}\!\left(\gamma_n\right)\,,
\end{split}\end{equation*}
which proves Lemma \eqref{lem:bhatt}.

\section{Proof of Theorem \ref{th:bstar}}\label{app:th3}
We will start from the Bhattacharyya coefficient and show that if, for any optimal interval $\mathcal{I}^{\star}_i$ the inequality in \eqref{eq:th2} holds, then the inequality in \eqref{eq:bplus} holds and through Lemma \ref{lem:Binf} we conclude the desired result. For notational simplicity, let
\begin{equation*}\begin{split}
u_i&\triangleq \sqrt{p_0(i)p_1(i)}\leq 1\,,\\
v_i&\triangleq \sqrt{p_0(0\vert i)p_1(0\vert i)}+\sqrt{p_0(1\vert i)p_1(1\vert i)}\,,\\
w_i&\triangleq \int_{\mathcal{I}^\star_i}\!\sqrt{f_0\left(x\vert i\right)f_1\left(x\vert i\right)}\,dx\,.
\end{split}\end{equation*}
It follows from \eqref{eq:bstardef} that
\begin{equation*}
b^\star_r=\sum_{i=1}^{2^r} u_i\,,
\end{equation*}
and from \eqref{eq:bhattinf} and \eqref{eq:binfty} that
\begin{equation*}\begin{split}
b_\infty&=\int_{\mathcal{X}}\!\sqrt{f_{X|H}(x|0) f_{X|H}(x|1) }\,dx\\
&=\sum_{i=1}^{2^r}\int_{\mathcal{I}^\star_i}\!\sqrt{f_{X|H}(x|0) f_{X|H}(x|1) }\,dx\\
&=\sum_{i=1}^{2^r}\sqrt{p_0\!\left(i\right)p_1\!\left(i\right)}\int_{\mathcal{I}^\star_i}\!\sqrt{f_0\!\left(x\vert i\right)f_1\!\left(x\vert i\right)}\,dx\\
&=\sum_{i=1}^{2^r}u_i w_i\,.
\end{split}\end{equation*}
Now let $b_{r+1}$ be the Bhattacharyya coefficient of the $r+1$ bit monotone quantizer resulting from dividing each interval $\mathcal{I}^\star_i$ in the optimum $r$ bit monotone quantizer by a threshold $\eta_i$ which satisfies \eqref{eq:th2}. Then
\begin{equation*}\begin{split}
b_{r+1}^2&=\left (\sum_{i=1}^{2^r}\sqrt{p_0\!\left(i,0\right)p_1\!\left(i,0\right)}+\sqrt{p_0\!\left(i,1\right)p_1\!\left(i,1\right)} \right)^2\\
&\stackrel{(a)}{=}\left (\sum_{i=1}^{2^r}u_iv_i\right)^2\\
&\stackrel{(b)}{\leq}\left (\sum_{i=1}^{2^r}u_i\sqrt{w_i}\right)^2\\
&\stackrel{}{=}\left (\sum_{i=1}^{2^r}\sqrt{u_i}\sqrt{u_iw_i}\right)^2\\
&\stackrel{(c)}{\leq}\left (\sum_{i=1}^{2^r}{u_i}\right)\left (\sum_{i=1}^{2^r}{u_iw_i}\right)\\
&=b^\star_r b_\infty\,.
\end{split}\end{equation*}
where $p_h\!\left(i,j\right)$ for $h,j \in \{ 0,1\}$ is defined in \eqref{eq:p_hij}, where $(a)$ follows as $p_h\!\left(i,j\right) = p_h(j\vert i)p_h(i)$, where
$(b)$ follows by \eqref{eq:th2}, and where $(c)$ follows by the Cauchy-Schwarz inequality.

\section{Proof of Lemma \ref{lem:Laplacian}}\label{app:lemm4}
According to our discussion in \ref{sec:lapla}, when $r \geq 1$ all the thresholds are in the interval $[-m,m]$, and consequently three different cases can happen for an arbitrary optimum interval $\mathcal{I}^\star_i\triangleq\left[a,c\right]$: (i) $-m\leq a< c \leq m$, (ii) $-m\leq a\leq  m$ and $c=\infty$, and (iii) $a=-\infty$ and $-m\leq c\leq  m$. In addition to these three cases, when $r=0$ we have $a=-\infty$ and $c=\infty$ as there are no thresholds when $r=0$. In what follows, we will be considering the different possible cases for an optimum interval and show that for each of them there is a threshold $\eta_i$ that satisfies the inequality \eqref{eq:th2}. As noted, this implies  that rate balancing is an optimal strategy when observation model at sensors is as \eqref{eq:Laplacase}. Because of symmetry, proving case (ii) will immediately result in case (iii) and therefore we only prove case (ii) explicitly. 

For a general $\mathcal{I}^\star_i=\left[a,c\right]$ and $h \in \{0,1\}$, we have
\begin{equation}\begin{split}
f_h\left(x\vert i\right)&=\frac{ f_{X\vert H}(x\vert h)}{\int_a^c f_{X\vert H}(x\vert h)\,dx}\,,\\
p_h\left(0\vert i\right)&={\int_a^{\eta_i}\! f_{h}(x\vert i)\,dx}\,,\\
p_h\left(1\vert i\right)&={\int_{\eta_i}^c \! f_{h}(x\vert i)\,dx}\,.
\label{eq:concatenatedprob}\end{split}\end{equation}

\subsubsection{$r>0$, and $\,-m\leq a< c\leq m$}
In this case, it follows straightforwardly from the definition in \eqref{eq:binfidef} and by direct integration that \[b_{\infty\vert i}=\frac{c-a}{\sqrt{e^{-a+c}-2+e^{-c+a}}}\,,\]
and for the explicit choice of $\eta_i=\frac{a+c}{2}$, it can be shown that
\[\left(b_{1\vert i}\right)^2=4\,\frac{e^{c/2-a/2}-2+e^{a/2-c/2}}{{e^{-a+c}-2+e^{-c+a}}}\,.\]
What remains to be shown is that for any $a< c$, it holds that
\begin{equation}
\left(b_{1\vert i}\right)^2\leq b_{\infty\vert i}\,.
\label{eq:inequalitybi}\end{equation}
Let $y\triangleq \frac{c-a}{2}$, where $y>0$. After some straightforward manipulations, it can be shown that proving \eqref{eq:inequalitybi} is equivalent to proving
\begin{equation*}
(2-y)e^y+(2+y)e^{-y}\leq4\,.
\end{equation*}
Note that $g(y)\triangleq(2-y)e^y+(2+y)e^{-y}$ is a monotone decreasing function of $y \geq 0$, which can be seen by noting that $g'(0) = 0$ and $g''(y) = -y(e^{y}-e^{-y}) < 0$ for $y > 0$. It follows that $g(y)$ is maximized at $y=0$, where $g(y=0)=4$. The inequality in \eqref{eq:inequalitybi} follows.
\subsubsection{$r>0$, and $-m \leq a\leq m, \,c=\infty$}
In this case we have \[b_{\infty\vert i}=\left(1+m-a\right)\,\frac{e^{\frac{a-m}{2}}}{\sqrt{2-e^{a-m}}}\,,\] and by setting $\eta_i=\frac{a+m}{2}$ we obtain \[\left(b_{1\vert i}\right)^2=\frac{e^\frac{a-m}{2}}{2-e^{a-m}}\left (\left(1-e^{\frac{a-m}{2}}\right)+\sqrt{{2-e^\frac{a-m}{2}}}\,\right )^2\,.\]
Define $y\triangleq e^{\frac{a-m}{2}}$, where $0\leq y\leq 1$ as $a \leq m$. Following \eqref{eq:inequalitybi} and after some manipulations our goal is to show
\begin{equation}
\left (1-y +\sqrt{2-y} \,\right)^2 \leq \left( 1-2\ln y\right )\sqrt{2-y^2}\,,
\label{eq:proofcase3cont}\end{equation}
for $0\leq y\leq 1$. In order to prove this inequality, we first define $\lambda(y)\triangleq \left (1-y +\sqrt{2-y} \,\right)^2$ and $\rho(y)\triangleq \left( 1-2\ln y\right )\sqrt{2-y^2}$. We will show that there is an auxiliary function $\mu(y)$ which satisfies $\lambda(y)\leq \mu(y)\leq \rho(y)$, when $0\leq y\leq 1$. We will repeatedly use Taylor's theorem \cite{Kap02}, which states that
\[\xi(y)=\xi(1)+(y-1)\xi'(1)+\frac{1}{2}(y-1)^2\xi''(1)+\frac{1}{6}(y-1)^3\xi'''(z)\]
for any continuous function $\xi(y)$ over $y \in [0,1]$, and for some $z \in [y,1]$ (where $z$ may depend on $y$ and $\xi$).

First let $\xi(y)=\lambda(y)$. It follows that
\begin{equation*}\begin{split}
\lambda(y)&=1-3(y-1)+2(y-1)^2+\frac{1}{6}(y-1)^3\lambda'''(z)\\
&={6-7y+2y^2}+{\frac{1}{6}(y-1)^3\lambda'''(z)}\\
&=\mu(y)+\epsilon_\lambda\,,
\end{split}\end{equation*}
where \[\mu(y)\triangleq 6-7y+2y^2\,,\] and \[\epsilon_\lambda\triangleq{\frac{1}{6}(y-1)^3\lambda'''(z)}\,.\] Now we show that $\epsilon_\lambda \leq 0$ and therefore that $\lambda(y)\leq \mu(y)$. To do so, it is needed to show that $\lambda'''(z) \geq 0$. We simply find
\[\lambda'''(z)=\frac{3(3-z)}{2(2-z)^{5/2}} \geq 0\]
for $0\leq z \leq 1$. Therefore $\lambda(y)\leq \mu(y)$.

Second, let $\xi(y)=\rho(y)$, then again using Taylor's theorem
\begin{equation*}\begin{split}
\rho(y)&=1-3(y-1)+2(y-1)^2+\frac{1}{6}(y-1)^3\rho'''(z)\\
&={6-7y+2y^2}+{\frac{1}{6}(y-1)^3\rho'''(z)}\\
&=\mu(y)+\epsilon_r\,,
\end{split}\end{equation*}
where \[\epsilon_\rho\triangleq \frac{1}{6}(y-1)^3\rho'''(z)\,.\]
Now we show that $\epsilon_\rho \geq 0$ and therefore that $\mu(y)\leq \rho(y)$ and the lemma is proved. To do so, it is needed to show $\rho'''(z) \leq 0$. Taking the third derivative of $\rho(z)$, i.e.,
\[\rho'''(z)=-2\,\frac{z^6+9z^4-6z^4\ln(z)-24z^2+16}{z^3(2-z^2)^{5/2}}\,,\]
implies that in order to get the outcome of interest we should show that for any $z\in[0,1]$
\[z^6+9z^4-6z^4\ln(z)-24z^2+16 \geq 0\,.\]
Using the inequality $\ln(z)\leq z-1$, for $0\leq z\leq 1$ we obtain
\begin{equation}\begin{split}
z^6+9z^4-6z^4&\ln(z)-24z^2+16 \\ &\geq  z^6-6z^5+15z^4-24z^2+16\\
& > z^6-6z^5+15z^4-24z^2+16 - (6-4z^3)\\
&= (z-1)^6 + (z-1)^2(9+24z) \,\geq 0\,.
\end{split}\end{equation}
This completes the proof of \eqref{eq:proofcase3cont}.
\subsubsection{$r=0$}
In this case $\mathcal{I}^\star_1=[-\infty,\infty]$ and our goal is to show that the Bhattacharyya distance of an optimal rate-one sensor is more than half of the Bhattacharyya distance contained in each observation (cf. \cite{Cham03}), or in terms of the Bhattacharyya coefficient, that $b_{1}^2\leq b_{\infty}$. It is straightforward to show that \[b_\infty=e^{-m}\left(m+1\right)\,,\] and by setting $\eta_1=0$ it follows that
\begin{equation*}\begin{split}
\left(b_1\right)^2&=e^{-m}\left(2-e^{-m}\right)\,.
\end{split}\end{equation*}
Then our goal is to show \[1-m-e^{-m}\leq 0\,,\] for $m\geq 0$. Defining $g(m)\triangleq 1-m-e^{-m}$, it can be seen that $g(m)$ is a monotone decreasing function of $m\geq 0$, and $g(0)=0$. This proves \eqref{eq:inequalitybi} for the final case and completes the proof of Lemma \ref{lem:Laplacian}.

\bibliographystyle{IEEEtran}
\bibliography{Ref}

\begin{thebibliography}{10}
\providecommand{\url}[1]{#1}
\csname url@samestyle\endcsname
\providecommand{\newblock}{\relax}
\providecommand{\bibinfo}[2]{#2}
\providecommand{\BIBentrySTDinterwordspacing}{\spaceskip=0pt\relax}
\providecommand{\BIBentryALTinterwordstretchfactor}{4}
\providecommand{\BIBentryALTinterwordspacing}{\spaceskip=\fontdimen2\font plus
\BIBentryALTinterwordstretchfactor\fontdimen3\font minus
  \fontdimen4\font\relax}
\providecommand{\BIBforeignlanguage}[2]{{%
\expandafter\ifx\csname l@#1\endcsname\relax
\typeout{** WARNING: IEEEtran.bst: No hyphenation pattern has been}%
\typeout{** loaded for the language `#1'. Using the pattern for}%
\typeout{** the default language instead.}%
\else
\language=\csname l@#1\endcsname
\fi
#2}}
\providecommand{\BIBdecl}{\relax}
\BIBdecl

\bibitem{Veer12}
V.~Veeravalli and P.~K. Varshney, ``Distributed inference in wireless sensor
  networks,'' \emph{Phil. Trans. A, Math. Phys. Eng. Sci.}, vol. 370, no. 1958,
  pp. 100--117, 2012.

\bibitem{Cham03}
J.-F. Chamberland and V.~Veeravalli, ``Decentralized detection in sensor
  networks,'' \emph{IEEE Trans. Signal Process.}, vol.~51, no.~2, pp. 407--416,
  Feb 2003.

\bibitem{Cham07}
------, ``Wireless sensors in distributed detection applications,'' \emph{IEEE
  Signal Process. Mag.}, vol.~24, no.~3, pp. 16--25, 2007.

\bibitem{li07}
W.~Li and H.~Dai, ``Distributed detection in wireless sensor networks using a
  multiple access channel,'' \emph{IEEE Trans. Signal Process.}, vol.~55,
  no.~3, pp. 822--833, 2007.

\bibitem{Varsh96}
P.~K. Varshney, \emph{Distributed Detection and Data Fusion}.\hskip 1em plus
  0.5em minus 0.4em\relax Springer-Verlag New York, Inc., 1996.

\bibitem{Berg09}
C.~R. Berger, M.~Guerriero, S.~Zhou, and P.~Willett, ``Pac vs. mac for
  decentralized detection using noncoherent modulation,'' \emph{IEEE Trans.
  Signal Process.}, vol.~57, no.~9, pp. 3562--3575, 2009.

\bibitem{Li11}
F.~Li, J.~S. Evans, and S.~Dey, ``Decision fusion over noncoherent fading
  multiaccess channels,'' \emph{IEEE Trans. Signal Process.}, vol.~59, no.~9,
  pp. 4367--4380, 2011.

\bibitem{Ciu13}
D.~Ciuonzo, G.~Romano, and P.~S. Rossi, ``Optimality of received energy in
  decision fusion over rayleigh fading diversity mac with non-identical
  sensors,'' \emph{IEEE Trans. Signal Process.}, vol.~61, no.~1, pp. 22--27,
  2013.

\bibitem{Poor77}
H.~V. Poor and J.~B. Thomas, ``Applications of {A}li-{S}ilvey distance measures
  in the design generalized quantizers for binary decision systems,''
  \emph{IEEE Trans. Commun.}, vol.~25, no.~9, pp. 893--900, 1977.

\bibitem{Lon90}
M.~Longo, T.~Lookabaugh, and R.~Gray, ``Quantization for decentralized
  hypothesis testing under communication constraints,'' \emph{IEEE Trans. Inf.
  Theory}, vol.~36, no.~2, pp. 241--255, Mar. 1990.

\bibitem{Swa93}
P.~Swaszek, ``On the performance of serial networks in distributed detection,''
  \emph{IEEE Trans. Aerosp. Electron. Syst.}, vol.~29, no.~1, pp. 254--260,
  1993.

\bibitem{Sun01}
W.~Shi, T.~Sun, and R.~Wesel, ``Quasi-convexity and optimal binary fusion for
  distributed detection with identical sensors in generalized {G}aussian
  noise,'' \emph{IIEEE Trans. Inf. Theory}, vol.~47, no.~1, pp. 446--450, 2001.

\bibitem{Ald04}
S.~A. Aldosari and J.~M. Moura, ``Detection in decentralized sensor networks,''
  in \emph{Proc. IEEE Int. Conf. Acoustics, Speech and Signal Processing
  (ICASSP)}, vol.~2, 2004, pp. ii--277.

\bibitem{Tsi88}
J.~N. Tsitsiklis, ``Decentralized detection by a large number of sensors,''
  \emph{Math. Contr., Signals, Syst.}, vol.~1, no.~2, pp. 167--182, 1988.

\bibitem{Cher52}
H.~Chernoff, ``A measure of asymptotic efficiency for tests of a hypothesis
  based on the sum of observations,'' \emph{Ann. Math. Stat.}, pp. 493--507,
  1952.

\bibitem{Tsi93}
J.~N. Tsitsiklis, ``Decentralized detection,'' \emph{Adv. Statist. Signal
  Process.}, vol.~2, no.~2, pp. 297--344, 1993.

\bibitem{Kai67}
T.~Kailath, ``The divergence and bhattacharyya distance measures in signal
  selection,'' \emph{IEEE Trans. Commun.}, vol.~15, no.~1, pp. 52--60, Feb.
  1967.

\bibitem{Boekee79}
D.~Boekee and J.~van~der Lubbe, ``Some aspects of error bounds in feature
  selection,'' \emph{Pattern Recognition}, vol.~11, no. 5--6, pp. 353--360,
  1979.

\bibitem{Tsi93Ext}
J.~N. Tsitsiklis, ``Extremal properties of likelihood-ratio quantizers,''
  \emph{IEEE Trans. Commun.}, vol.~41, no.~4, pp. 550--558, 1993.

\bibitem{Mur03}
K.~Murota, \emph{Discrete convex analysis}.\hskip 1em plus 0.5em minus
  0.4em\relax SIAM, 2003, vol.~10.

\bibitem{Mar10}
A.~W. Marshall, I.~Olkin, and B.~C. Arnold, \emph{Inequalities: Theory of
  Majorization and Its Applications: Theory of Majorization and Its
  Applications}.\hskip 1em plus 0.5em minus 0.4em\relax Springer, 2010.

\bibitem{Poor88}
H.~V. Poor, ``Fine quantization in signal detection and estimation,''
  \emph{IEEE Trans. Inf. Theory}, vol.~34, no.~5, pp. 960--972, 1988.

\bibitem{Ben89}
G.~R. Benitz and J.~A. Bucklew, ``Asymptotically optimal quantizers for
  detection of iid data,'' \emph{IEEE Trans. Inf. Theory}, vol.~35, no.~2, pp.
  316--325, 1989.

\bibitem{Alla15}
A.~Tarighati and J.~Jald{\'e}n, ``Rate allocation for decentralized detection
  in wireless sensor networks,'' in \emph{Proc. 16th IEEE Int. Workshop Signal
  Process. Advances in Wireless Commun. (SPAWC)}, June 2015, pp. 341--345.

\bibitem{Alla14}
------, ``Bayesian design of decentralized hypothesis testing under
  communication constraints,'' in \emph{Proc. IEEE Int. Conf. Acoustics, Speech
  and Signal Processing (ICASSP)}, May 2014, pp. 7624--7628.

\bibitem{Kap02}
K.~Wilfred, \emph{Advanced calculus}.\hskip 1em plus 0.5em minus 0.4em\relax
  Addison-Wesley Longman, Boston, 2002.

\end{thebibliography}

\end{document}